\documentclass[10pt,aps,prd,showpacs,twocolumn,superscriptaddress]{revtex4-2}
\usepackage[usenames,dvipsnames]{color}
\usepackage{amsfonts}
\usepackage{amssymb}
\usepackage{amsmath}
\usepackage{graphicx}
\usepackage[colorlinks]{hyperref}
\usepackage[normalem]{ulem}
\usepackage{mathtools}

\begin{document}

\title{Neutron stars in the symmetron model}

\author{Bernardo F.\ de Aguiar}
\email{bernardo\_aguiar@id.uff.br }
\affiliation{CBPF - Centro Brasileiro de Pesquisas F\'isicas, Rio de Janeiro, 22290-180, Brazil.}

\author{Raissa F.\ P.\ Mendes}
\email{rfpmendes@id.uff.br}
\affiliation{Instituto de F\'isica, Universidade Federal Fluminense, Niter\'oi, 
Rio de Janeiro, 24210-346, Brazil.}

\author{F.\ T.\ Falciano}
\email{ftovar@cbpf.br}
\affiliation{CBPF - Centro Brasileiro de Pesquisas F\'isicas, Rio de Janeiro, 22290-180, Brazil.}
\affiliation{PPGCosmo, CCE - Federal University of Esp\'\i rito Santo, zip 29075-910, Vit\'oria, ES, Brazil.}

\date{\today}

\begin{abstract}
Screening mechanisms are often deployed by dark energy models in order to conceal the effects of their new degrees of freedom from the scrutiny of terrestrial and solar system experiments. 
However, extreme properties of nuclear matter may lead to a partial failure of screening mechanisms inside the most massive neutron stars observed in Nature, opening up the possibility of probing these theories with neutron star observations. In this work we explore equilibrium and stability properties of neutron stars in two variants of the symmetron model. 
We show that around sufficiently compact neutron stars, the symmetron is amplified with respect to its background, cosmological value by several orders of magnitude, and that properties of such unscreened stars are sensitive to corrections to the leading linear coupling between the symmetron and matter. 
\end{abstract}


\maketitle

\section{Introduction}

Due to their high compactness, neutron stars (NSs) offer a unique environment to probe the strong-field regime of Einstein's general relativity (GR) and to constrain possible modifications to it. Moreover, their core is characterized by extreme densities and pressures, which may lead to additional, matter-induced phenomenology in alternative theories of gravity, as compared, e.g., to the case of black holes. Exploring these effects becomes particularly relevant with the increasing accuracy of measurements of NS properties, inferred both through their electromagnetic and gravitational-wave emission \cite{Ozel2016,TheLIGOScientificCollaboration2018,Abbott2021,theligoscientificcollaboration2021gwtc3}.

In this work we focus on scalar extensions of GR where, in additional to the usual spin-2 field, gravity is mediated by a self-interacting scalar degree of freedom, characterized by a potential $V(\phi)$ and an effective coupling $\alpha(\phi) \equiv (\ln A(\phi))_{,\phi}$ to matter \cite{Pietroni2005PRD,Olive2008PRD,Khoury2010,Damour1992,Damour1994NuPB}.
Scalar-tensor theories of this kind offer a suitable framework for cosmology \cite{baker2019,Koyama2016,Bull2016}, since a judicious choice of $V(\phi)$ and $A(\phi)$ may lead to a model which behaves as dark energy \cite{Clifton2012, Joyce2015} at cosmological scales but still reproduces the successes of general relativity in explaining solar system and other observational data \cite{Will2006,Sakstein2020}. 
Typically, this is accomplished through a suppression, or {\it screening}, of scalar field effects at solar system (or galaxy) scales, which exploits the fact that the scalar field dynamics is governed by a density-dependent effective potential, $V_\textrm{eff}(\phi) \equiv V(\phi) - T \ln A(\phi)$, where $T$ is the trace of the energy-momentum tensor of matter fields (for a non-relativistic fluid, $T \approx -\rho$, where $\rho$ is the fluid rest-mass density).

Perhaps the most well-known example of screening (of the type described above) is that implemented in the chameleon model \cite{Khoury2004,Khoury2004a}. By combining a power-law potential and a constant effective coupling to matter, the chameleon field is endowed with a density-dependent effective mass. Thus, the field becomes massive and short-ranged in high density environments (such as the solar system), but light and long-ranged at cosmological scales, possibly behaving as dark energy. 

Another example of screening mechanism, which is the focus of the present work, occurs in the symmetron model \cite{Hinterbichler2010,Hinterbichler2011} (see also Refs.~\cite{Brax2012,Brax201238,Burrage2016PRL} for generalizations). Screening in this model relies on the spontaneous breaking of a $\mathbb{Z}_2$ symmetry at low densities, and its restoration in high density environments. In the latter case, the scalar coupling to matter is suppressed, since it is proportional to the local vacuum expectation value of the scalar field, which vanishes in the symmetric phase. Originally, the symmetron was introduced as an alternative model to explain the late-time accelerated expansion of the universe \cite{Hinterbichler2010,Hinterbichler2011}, and had its consequences explored in the context of linear and nonlinear structure formation~\cite{Brax2011,Brax2012JCAP,Davis2012,Llinares2013,Llinares2014, Mota2018IJMPD}, as well as for dark matter halo properties \cite{Clampitt2012,Taddei2014PRD, Contigiani2019PRD}. More recently, it has also found applications as a model for dark matter \cite{Burrage2017,Ohare2018PRD,Burrage2019}.

In order to probe chameleon/symmetron effects, one typically seeks for low-density, possibly unscreened environments, like those found under special laboratory conditions (see, e.g.,~Refs.~\cite{Upadhye2013,Burrage2016b,Cronenberg2018NatPh,Chiow2020PRvD,Elder2020} and \cite{Burrage2018} for a review). One the other hand, one might naively expect that NSs would be completely self-screened once the model parameters have been tuned to suppress dark energy effects in the solar system, as the NS mean density is orders of magnitude larger. Indeed, this is typically the case, as found in initial investigations in the chameleon and environmentally dependent dilaton 
models \cite{Babichev2010,Brax2017}. 

Interestingly, however, it has been pointed out that a partial failure of screening mechanisms may occur at the cores of the most massive, most compact NSs found in Nature \cite{Aguiar2020}. For a perfect fluid, the trace of the energy-momentum tensor, to which the scalar field couples, is given by $T = 3p - \epsilon$, where $p$ is the pressure and $\epsilon$ is the energy density as measured in the fluid rest frame. If the nuclear equation of state is such that a pressure-dominated phase, with $p > \epsilon/3$, occurs in the core of a NS \cite{Podkowka2018,Saes2021}, then $T$ changes sign, resulting in a partial descreening of the stellar interior. This effect was analyzed in Ref.~\cite{Aguiar2020} for the chameleon and dilaton models, with a further, in-depth exploration of the chameleon model presented in Ref.~\cite{Dima2021}.

The aim of the present work is to extend the analysis of Ref.~\cite{Aguiar2020} for the symmetron model, by investigating equilibrium and stability properties of symmetron neutron stars (SNSs), with special attention to those with pressure-dominated cores.
This work is organized as follows. We begin, in Sec.~\ref{sec:framework}, by defining the symmetron model, discussing the symmetron screening mechanism, and reviewing the background cosmological evolution in the model. We then proceed, in Sec.~\ref{sec:SNS}, to describe our main results on the equilibrium and stability properties of SNSs. Section \ref{sec:conclusion} gathers our main conclusions. In what follows, $M_\text{Pl} = \sqrt{\hbar c/8\pi G}$ denotes the (reduced) Planck mass, and we use units such that $c = \hbar = 1$.

\section{Symmetron model} \label{sec:framework}

\subsection{Field equations}

The symmetron model belongs to a class of scalar-tensor theories described by the following action functional \cite{Damour1992,Brax2012},
\begin{align} \label{eq:generalaction}
S & =\int d^4 x \sqrt{-g}
\left[ \frac{1}{2} M_\text{Pl}^2 \mathcal{R} - \frac{1}{2} g^{\mu \nu} \nabla_{\mu} \phi \nabla_{\nu} \phi - V(\phi) \right]
\nonumber \\
& + S_{m}[\Psi_m ; A(\phi)^2 g_{\mu\nu}],
\end{align}
where $S_m$ denotes the action for matter fields $\Psi_m$, which couple universally to the scalar field $\phi$ through the conformally rescaled (Jordan-frame) metric $\tilde{g}_{\mu\nu} \equiv A(\phi)^2 g_{\mu\nu}$. Variation of Eq.~(\ref{eq:generalaction}) with respect to the (Einstein-frame) metric $g_{\mu\nu}$ and scalar field $\phi$ yields the following field equations:
\begin{align} 
&G_{\mu \nu} = M_\text{Pl}^{-2} \left[ T_{\mu \nu } + \nabla_{\mu} \phi\nabla_{\nu} \phi - g_{\mu \nu} \left (\frac{1}{2}\nabla^{\beta}\phi\nabla_{\beta}\phi + V(\phi) \right ) \right], \label{eq:field1}\\ 
&\nabla^\mu \nabla_\mu \phi = \frac{d V }{d \phi} - \alpha(\phi) T,\label{eq:field2}
\end{align}
where 
\begin{equation}
    \alpha(\phi) \equiv \frac{d \ln A}{d\phi},
\end{equation}
and $T \equiv g^{\mu\nu} T_{\mu\nu}$ is the trace of the energy-momentum tensor of matter fields, $T_{\mu\nu} \equiv - 2 (-g)^{-1/2} \delta S_m/\delta g^{\mu\nu}$, which obeys
\begin{equation}\label{eq:eqmotion}
\nabla^\nu T_{\mu\nu} = \alpha(\phi) T \nabla_\mu \phi.
\end{equation}
It is often convenient to introduce the Jordan-frame energy-momentum tensor, $\tilde{T}_{\mu\nu} \equiv - 2 (-\tilde{g})^{-1/2} \delta S_m/\delta \tilde{g}^{\mu\nu} = A(\phi)^{-2} T_{\mu\nu}$, since this object is covariantly conserved in the sense that $\tilde{\nabla}^\nu \tilde{T}_{\mu\nu} = 0$, where quantities with tildes are constructed from the Jordan-frame metric.

More specifically, the symmetron model presented in Ref.~\cite{Hinterbichler2010} is characterized by a quartic, symmetry breaking potential,
\begin{equation}\label{eq:symmetronV}
V(\phi) = - \frac{1}{2} \mu^2 \phi^2 + \frac{\lambda}{4} \phi^4,
\end{equation}
and a conformal factor that respects its reflection symmetry under $\phi \to -\phi$. The simplest model therefore features a quadratic conformal factor,
\begin{equation}\label{eq:symmetronAq}
    A_\text{q}(\phi) = 1 + \frac{\phi^2}{2 M_s^2},
\end{equation}
where the dimension-full constant $M_s$ can be thought of as a cutoff scale below which corrections to (\ref{eq:symmetronAq}) can be safely ignored.
However, terms of order $O(\phi^4/M_s^4)$ or higher must be considered if the evolution drives the scalar field close to the cutoff scale. Interestingly, we will see that this can be the case for SNSs. Therefore, in our analysis, we will also consider the alternative, ``regularized'' variant
\begin{equation}\label{eq:symmetronAr}
    A_\text{r}(\phi) = 1 + \frac{\phi^2}{2 M_s^2 + M_\text{Pl} |\phi|},
\end{equation}
introduced in Ref.~\cite{Hinterbichler2011}, which serves the purpose of attenuating the scalar-mediated force for $\phi$ close to the cutoff scale $M_s$. 

\subsection{Screening}

As clear from Eq.~(\ref{eq:eqmotion}), in the symmetron model free particles do not follow geodesics of the Einstein-frame metric $g_{\mu\nu}$, but follow forced trajectories instead, with acceleration given by
\begin{equation} \label{eq:acceleration}
    a^{\mu} \equiv u^\nu \nabla_\nu u^\mu = - P^{\mu\nu} \partial_\nu \ln A,
\end{equation}
where $P^{\mu\nu} \equiv g^{\mu\nu} + u^\mu u^\nu$ projects onto the subspace orthogonal to the particle's four-velocity. If one expands the scalar field around its vacuum expectation value ($\phi_0$), $\phi = \phi_0 + \delta \phi$, then to leading order in the perturbation $\delta \phi$ the force per unit mass in Eq.~(\ref{eq:acceleration}) becomes
\begin{equation} \label{eq:symmetronforce}
    \vec{f}_\text{sym} = - \vec{\nabla} \ln A = - \alpha(\phi) \vec{\nabla} \phi 
    \approx \left(\frac{\phi_0}{M_s}\right) \vec{\nabla} \left(\frac{\delta \phi}{M_s}\right)\ .
\end{equation}
Thus, the coupling between matter and scalar field perturbations is proportional to $\phi_0$. This, in turn, depends on properties of the local matter environment, as follows.

From Eq.~(\ref{eq:field2}) we see that the scalar field responds to the effective potential
\begin{equation}\label{eq:Veff}
V_\textrm{eff}(\phi) \equiv V(\phi) - T \ln A(\phi),
\end{equation}
so that a constant solution $\phi = \phi_0$ must obey $d V_\textrm{eff} / d\phi |_{\phi_0} = 0$. After substituting Eqs.~(\ref{eq:symmetronV}) and (\ref{eq:symmetronAq}), and taking into account only the leading order contribution from $A(\phi)$, Eq.~(\ref{eq:Veff}) can be written as
\begin{equation}
V_\textrm{eff}(\phi) \approx \frac{1}{2} \left (- T - \mu^2 M_s^2 \right) \frac{\phi^2}{M_s^2} + \frac{1}{4} \lambda \phi^4.
\end{equation}
The sign of the quadratic term of the effective potential is seen to depend on the local matter content, through the trace of the energy-momentum tensor. If densities are sufficiently large ($\rho > \rho_* \equiv \mu^2 M_s^2$) and matter is non-relativistic, $T \approx - \rho < 0$,
the effective potential has a minimum at $\phi_0 = 0$, in which case the coupling to matter vanishes identically [cf.~Eq.~(\ref{eq:symmetronforce})]. 
On the other hand, in rarefied environments ($\rho < \rho_*$) the $\mathbb{Z}_2$ symmetry is broken as the field tends to settle at one of the nontrivial minima of the effective potential (at $\phi_0 = \pm \mu / \sqrt{\lambda}$ for $\rho = 0$). In this case, symmetron perturbations couple to matter with strength $\phi_0/M_s^2 \approx \mu /(M_s^2 \sqrt{\lambda})$, and can have a non-negligible impact on the cosmological evolution. Thus, the main ingredients of the symmetron screening mechanism are the restoration of the $\mathbb{Z}_2$ symmetry in high density environments, together with a coupling to matter that depends on the symmetron vacuum expectation value.

The symmetron model discussed in this work is characterized by three parameters, $\mu$, $M_s$, $\lambda$, the magnitude of which is guided by the intended applications, and restricted by observations. In particular, for the symmetron to provide a viable model for dark energy, it must become tachyonic around the current cosmic density, which means that the critical density $\rho_* = \mu^2 M_s^2$ for symmetry breaking must be of the order of the current cosmic density,
\begin{equation}\label{eq:muM}
    H_0^2 M_\text{Pl}^2 \approx \mu^2 M_s^2,
\end{equation}
where $H_0$ is the Hubble parameter (The background cosmology in the symmetron model will be revisited in Sec.~\ref{sec:cosmology}.). Additionally, for the symmetron field to drive cosmic expansion, it must mediate a force comparable to gravity:
\begin{equation}\label{eq:lambdaM}
    \frac{\mu}{M_s^2 \sqrt{\lambda}} \approx \frac{1}{M_\text{Pl}}.
\end{equation}
Conditions (\ref{eq:muM}) and (\ref{eq:lambdaM}) tie together the model constants, leaving only one independent parameter, which can be taken as the cutoff scale $M_s$. This, in turn, can be constrained by local experiments and astrophysical observations \cite{Hinterbichler2010,Jain2013,baker2019,Sakstein2020}. In particular, requiring the Milky Way to be screened enforces 
\begin{equation} \label{eq:constraintMs}
    M_s \lesssim 10^{-3} M_\text{Pl}.
\end{equation}
In this case, the range of the symmetron-mediated force in vacuum is of the order of $\mu^{-1} \lesssim 10^{-3} H_0^{-1} \sim $ 1 Mpc. 

It is worth mentioning that for other applications -- e.g.~considering the symmetron as a model for dark matter \cite{Burrage2017,Ohare2018PRD,Burrage2019} -- one does not need to impose conditions (\ref{eq:muM}) and (\ref{eq:lambdaM}), leaving a larger space of parameters to be contrasted with observations (see \cite{Burrage2018} for a review). For instance, Ref.~\cite{Upadhye2013} explores constraints to $\lambda$ from torsion-pendulum experiments by fixing $\mu$ according to the dark energy scale and $M_s \sim 1$ TeV, just beyond probed Standard Model energies.
For definiteness, however, we will consider the model parameters to be tied together as in Eqs.~(\ref{eq:muM}) and (\ref{eq:lambdaM}) in the present work.

\subsection{Cosmology} \label{sec:cosmology}

In this section we revisit the main aspects of the cosmological solution of a Friedmann-Lemaitre-Robertson-Walker (FLRW) universe populated with a collection of fluid species and a symmetron field $\phi = \phi(t)$. The Einstein-frame metric assumes the (spatially flat) FLRW form
\begin{equation}
    ds^2 = - dt^2 + a(t)^2 (dx^2 + dy^2 + dz^2).
\end{equation}
In this coordinate system the energy-momentum tensor for matter fields has components $T^\mu_{\;\;\; \nu} = \sum_i \text{diag}(-\epsilon_i, p_i, p_i, p_i) $, with pressure and energy density assumed to be related by a constant equation of state, $p_i = w_i \epsilon_i$, for each fluid species $i$. Assuming that Eq.~(\ref{eq:eqmotion}) holds for each fluid species separately, one finds that 
\begin{equation}
    \epsilon_i = \epsilon_{i,0} \, a^{-3 (1+w_i)} A^{1 - 3w_i},
\end{equation}
where $\epsilon_{i,0}$ are constants, in terms of which one can define the fractional abundances $\Omega_i \equiv \epsilon_{i,0} / (3 H_0^2 M_\text{Pl}^2)$.

The dynamical equations (\ref{eq:field1}) and (\ref{eq:field2}) imply
\begin{align}
&3 M_\text{Pl}^2 H^2 = \frac{1}{2} \dot{\phi}^2 + V(\phi) + \sum_i \epsilon_i, \label{eq:cosmo1}\\
&\ddot{\phi} + 3H\dot{\phi} + \frac{d V}{d \phi} + \alpha \sum_i (1-3w_i)\epsilon_i=0, \label{eq:cosmo2}
\end{align}
where dots stand for time derivatives with respect to the Einstein-frame cosmic time $t$, and $H \equiv \dot{a}/a$.

Assuming that the symmetron field exits inflation with a value $\phi_i \lesssim M_s$, one finds that it evolves during the radiation and matter dominated eras as follows (see Ref.~\cite{Hinterbichler2011} for details). Initially, Hubble friction [encapsulated by the second term of Eq.~(\ref{eq:cosmo2})] dominates, and the symmetron remains frozen at $\phi_i$ until $a \approx a_{\rm eq} M_s^2/(3 M_{\rm Pl}^2)$, where $a_{\rm eq}$ denotes the scale factor at matter-radiation equality. At this point---which is well before matter-radiation equality since $M_s \ll  M_{\rm Pl}$---, the coupling to matter overcomes friction, and the symmetron begins to perform damped oscillations around the minimum of the effective potential at $\phi_0 = 0$. Around $a=0.5$, its amplitude has decayed to a value $\sim 10^{-3} (M/M_{\rm Pl})^{3/2} \phi_i$. However, as expansion proceeds and matter density drops below the critical value $\rho_*  = \mu^2 M_s^2$, a phase transition takes place and $\phi_0 $ moves to one of the symmetry-breaking minima of the effective potential. 

\begin{figure}[tbh]
\centering
\includegraphics[width=8cm]{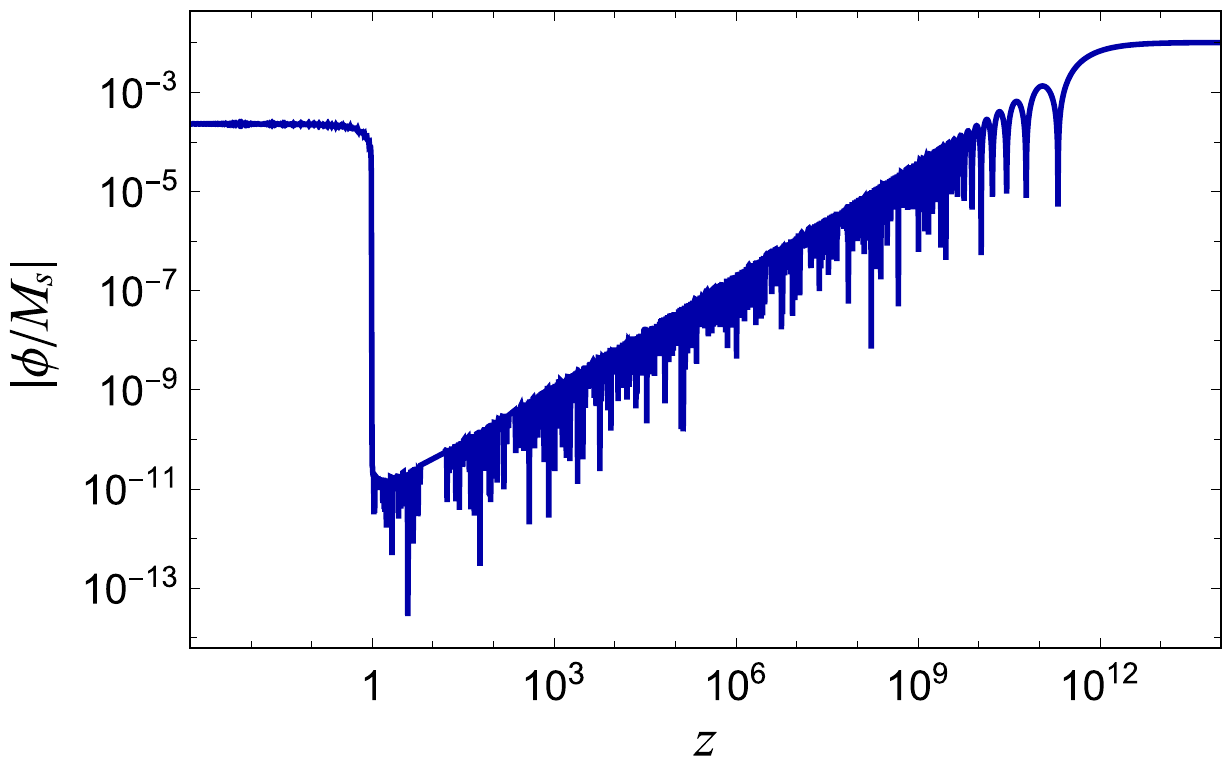}
\caption{Cosmological evolution of the symmetron field as a function of redshift in a universe field with radiation ($w_\gamma = 1/3$, $\Omega_\gamma \sim 10^{-4}$), pressureless matter ($w_m = 0$, $\Omega_m$ = 0.25) and a cosmological constant ($w_\Lambda = -1$, $\Omega_\Lambda \approx 1-\Omega_m$), for the quadratic variant of the symmetron model, with $M_s = 10^{-4} M_\text{Pl}$, $\mu = \sqrt{15} H_0 M_\text{Pl}/M_s$, and $\lambda = \mu^2 M_\text{Pl}^2 /M_s^2$. For this choice of parameters, the phase transition occurs at $z \approx 1$. }
\label{fig:cosmo}
\end{figure}

The cosmological evolution of the symmetron field, as per Eqs.~(\ref{eq:cosmo1}) and (\ref{eq:cosmo2}), is shown in Fig.~\ref{fig:cosmo}, where the phases described above are clearly identifiable. As pointed out in Ref.~\cite{Hinterbichler2011}, for the simplest choice of a quartic potential, as in Eq.~(\ref{eq:symmetronV}), and taking conditions (\ref{eq:muM})-(\ref{eq:constraintMs}) to be valid, the symmetron potential energy is not enough to drive cosmic acceleration. Thus, in order to reproduce $\Lambda$CDM expansion history in this model, a cosmological constant must be included, e.g., in the form of a nondynamical constant $V_0$ added to the potential or as fluid species with $w_\Lambda = -1$. In constructing Fig.~\ref{fig:cosmo} we have adopted the latter, letting $\Omega_\Lambda \approx 1 - \Omega_m$, and $\Omega_m = 0.25$. The symmetron was initialized to $\phi_i = 10^{-2} M_s$, and model parameters were adjusted so that the phase transition occured at redshift $z \approx 1$. One can see that the symmetron field is successfully brought close to the symmetry-restoring point (at $\phi_0 = 0$), before a phase transition takes place at the present age, amplifying the symmetron to values close to the symmetry breaking point. 
However, in this case the fractional energy density in the symmetron field still remains small, which challenges its viability to drive the late time acceleration of the universe. Notwithstanding, our discussion also shows that the symmetron model provides a suitable screening mechanism for astrophysical scales. From now on, we will be interested in the symmetron field configurations inside dense neutron stars.

\section{Equilibrium and stability properties of symmetron neutron stars} \label{sec:SNS}

\subsection{Set-up} \label{sec:setup}

In order to determine the structure of symmetron neutron stars (SNSs), we approximate the spacetime to be static and spherically symmetric, with line element
\begin{equation}\label{eq:lineelement1}
ds^2 = - e^{2 \nu(r)} dt^2 + e^{2\lambda(r)} + r^2 (d\theta^2 + \sin^2 \theta d\varphi^2).
\end{equation}
The NS is modeled as a perfect fluid, with energy-momentum tensor
\begin{equation} \label{eq:perfect_fluid}
    T^{\mu\nu} = (\epsilon + p) u^\mu u^\nu + p g^{\mu\nu},
\end{equation}
where $u^\mu$ is the four-velocity of fluid elements, and $\epsilon$ and $p$ are the energy density and pressure as measured by comoving observers. We further define the Jordan-frame pressure and energy-density, $\tilde{p} = A(\phi)^{-4} p$ and $\tilde{\epsilon} = A(\phi)^{-4} \epsilon$, in terms of which we specify the equation of state (EOS). We consider a barotropic EOS relating pressure and number density ($\tilde{n}$): $\tilde{p} = \tilde{p}(\tilde{n})$. In turn, the energy density is obtained by the first law of thermodynamics, $d (\tilde{\epsilon}/\tilde{n}) = -\tilde{p} d(1/\tilde{n})$, assumed to hold in the Jordan frame.
Specifically, in this work we adopt the ENG EOS \cite{Engvik1995}, in a piecewise-polytropic parametrization \cite{Read2009}.

With the assumptions above, one can derive the following set of structure equations from Eqs.~(\ref{eq:field1}) and (\ref{eq:field2}):
\begin{align}
&\frac{dm}{dr} = 4\pi r^2 \left[ A^4 \tilde{\epsilon} + \frac{1}{2} e^{-2\lambda} \psi^2 + V \right],
\label{eq:eqmb}
\\
&\frac{d\nu}{d r} = r e^{2\lambda} \left[ \frac{m}{r^3} + 4\pi A^4 \tilde{p} + 2\pi e^{-2\lambda} \psi^2 - 4\pi V \right], \label{eq:eqnub}
\\
&\frac{d\tilde{p}}{d r} = - (\tilde{p}+\tilde{\epsilon}) \frac{d}{dr} \left( \nu + \ln A \right),
\label{eq:eqpb}
\\
& \frac{d\phi}{dr} = \psi,\\
&\frac{d}{dr}\! \left( r^2 e^{\nu - \lambda} \psi \right) = r^2 e^{\nu + \lambda} \left[ \frac{d V}{d \phi} -  A^3 \frac{d A}{d\phi} (3\tilde{p} - \tilde{\epsilon})\right].
\label{eq:eqphib}
\end{align}
Here, the mass aspect function $m(r)$ is defined through $m(r) \equiv (r/2) (1-e^{-2\lambda(r)})$. For simplicity, vacuum is assumed outside of the star.

The system (\ref{eq:eqmb})-(\ref{eq:eqphib}), supplemented by the EOS relating pressure and energy density, can be solved by standard methods, with the following boundary conditions: $m(0) = 0$, so that the solution is regular at $r=0$; $\tilde{p}(R) = 0$, which defines the (Einstein-frame) stellar radius $R$; $\phi(r) \to \phi_0$ for $r \gg R$, where $\phi_0 = M_s^2/M_\text{Pl}$ is the (positive, for definiteness) symmetry breaking minimum of the effective potential (\ref{eq:Veff}), and $\nu(r) \to - (1/2) \ln [1-2m(r)/r] $ for $r\gg R$, such that the spacetime becomes Schwarzschild-de Sitter far away from the star.
The total mass satisfies $M \approx m(r) - \frac{4\pi}{3} r^3 V(\phi_0)$ for $r\gg R$; which differs only slightly from $m(R)$ in the models we consider. Results for equilibrium properties of SNSs will be discussed in Sec.~\ref{sec:results} below.

A fundamental additional step will be to establish whether the equilibrium solutions we construct are stable under linear radial perturbations. For that purpose, we begin by promoting $\nu$ and $\lambda$ in Eq.~(\ref{eq:lineelement1}) to functions of $(t,r)$, such that $\nu(t,r) = \nu_0 (r) + \delta \nu (t,r)$ and $\lambda(t,r) = \lambda_0 (r) + \delta \lambda (t,r)$, with $\nu_0(r)$ and $\lambda_0(r)$ denoting background quantities, and similarly for the scalar field, pressure and energy density. 
The perturbed fluid four-velocity
\begin{equation}
u^\mu (t,r) = e^{-\nu_0} (1-\delta \nu, d\xi/dt, 0, 0)
\end{equation}
is written in terms of $\xi(t,r)$, the radial Lagrangian displacement of a given fluid element.

The perturbed configuration is completely specified by six functions, $\delta \nu$, $\delta \lambda$, $\delta \phi$, $\xi$, $\delta \tilde{p}$, and $\delta \tilde{\epsilon}$. In Ref.~\cite{Aguiar2020}, it was shown that these functions can be written in terms of $\xi$ and $\delta \phi$, which obey a set of two coupled homogeneous second order partial differential equations. These master equations were derived under the assumption that both the perturbed and unperturbed configurations obey the same (cold) EOS. Assuming a harmonic time dependence of $\xi$ and $\delta \phi$, 
\begin{equation}\label{eq:harmonict}
\xi (t,r) = \xi (r) e^{i\omega t}, \qquad
\delta \phi (t,r) = \delta \phi (r) e^{i \omega t},
\end{equation}
with $\omega \in \mathbb{C}$, the master equations have the schematic form 
\begin{equation} \label{eq:diffall}
\frac{d {\bf x} (r)}{dr} = {\bf M}(r) {\bf x}(r),
\end{equation}
where ${\bf x} (r) = (\xi, \xi', \delta \phi, \delta \phi')^T$ (with a prime denoting a radial derivative) and ${\bf M}(r)$ is a $4\times 4$ matrix function of background quantities alone \footnote{Explicit expressions for ${\bf M}$ and ${\bf F}$ are available 
at \url{https://bitbucket.org/raissafpmendes/symmetron_neutron_stars.git}.}.

The boundary conditions are the following. Regularity at $r=0$ is ensured by taking $\xi(0) = 0$ and $\delta \phi' (0) = 0$, while regularity at $r = R$ gives rise to an additional requirement of the form ${\bf F}(R)^T {\bf x} (R) = 0$, where ${\bf F}(R)$ is a vector constructed from background quantities. Finally, since we will be looking for unstable modes, for which $\omega^2 <0$, we demand that $\delta \phi(r) \to 0$ for $r \gg R$. Since the system (\ref{eq:diffall}) is homogeneous, there is an overall normalization freedom (${\bf x} \to C {\bf x}$, with $C$ a constant), and the system is overdetermined by the four boundary conditions above. As a consequence, solutions are to be found only for a discrete (possibly empty) set of values for $\omega$. These are sought numerically through a shooting procedure, as described in more detail in Ref.~\cite{Aguiar2020}.

\subsection{Symmetron neutron stars} \label{sec:results}

In the Newtonian context, a simple criterion for an astrophysical body to be screened~\cite{Hinterbichler2010} is the surface Newtonian potential $\Phi_N$ to be much larger than the ratio $M_s^2/M_\text{Pl}^2$, i.e. the parameter
\begin{equation} \label{eq:alpha}
    \Upsilon \equiv 6 \Phi_N \frac{M_\text{Pl}^2}{M_s^2} \gg 1\ .
\end{equation}
Indeed, the parameter $\Upsilon$ determines to what extent the thin shell mechanism operates inside that object, with the thickness of the thin shell scaling as $\Delta R \sim \Upsilon^{-1} R$ \cite{Hinterbichler2010}. Consequently, $\Upsilon^{-1}$ also determines the ratio of the scalar-mediated force to gravity.

\begin{figure}[thb]
\centering
\includegraphics[width=8cm]{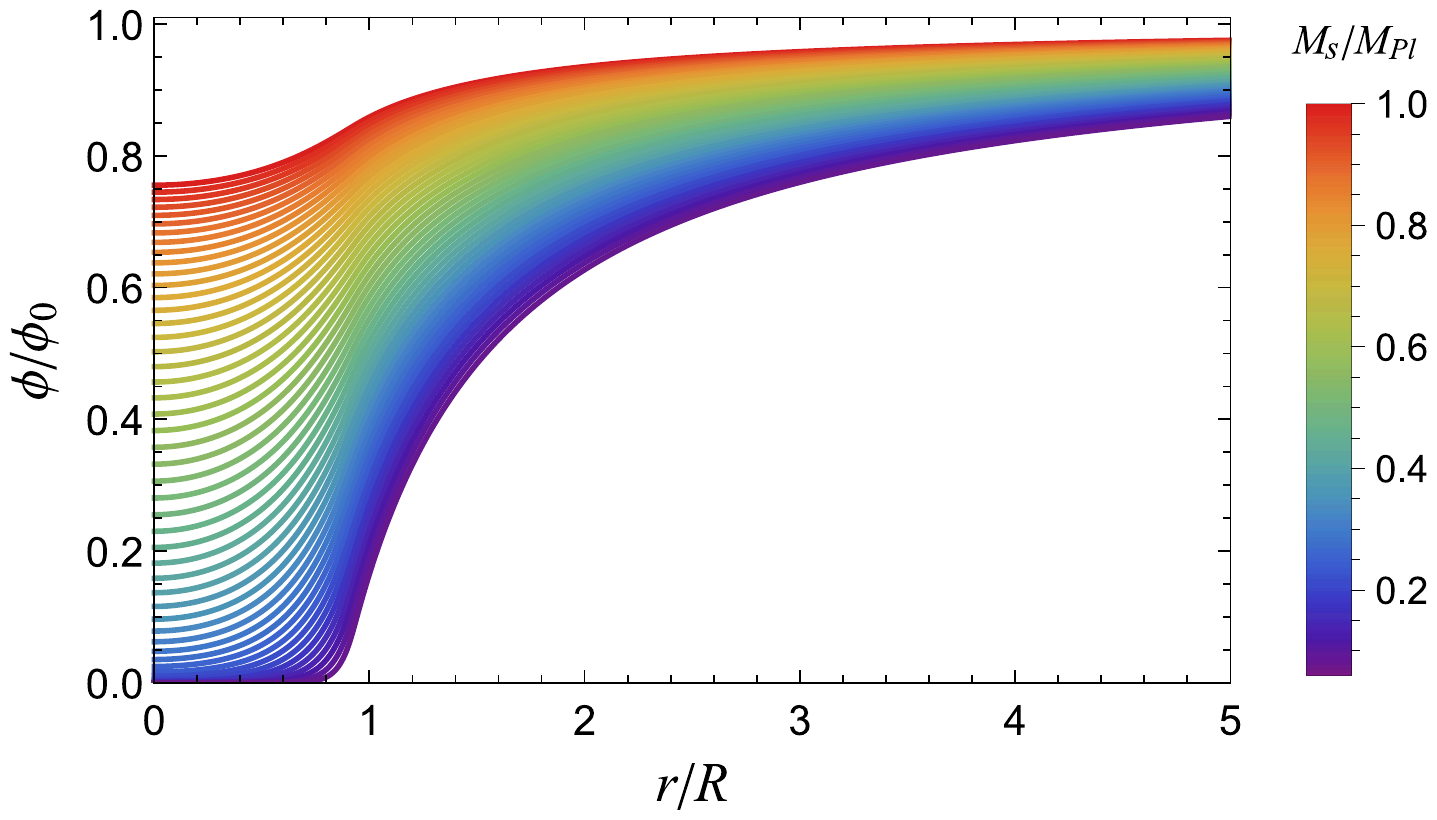}
\caption{Scalar field profile, rescaled by its asymptotic value ($\phi_0 = M_s^2/M_\text{Pl}$), as a function of the radial coordinate, for values of $M_s/M_\text{Pl}$ ranging from 0.06 to 1. In all cases, the central number density of the SNS was fixed to $0.476$ fm$^{-3}$, which, in GR, yields a $1.4 M_\odot$ equilibrium model for the ENG EOS used in this work. The masses of SNSs increase monotonically from $0.595 M_\odot$ when $M_s = M_\text{Pl}$ to $1.3999 M_\odot$ when $M_s = 0.06 M_\text{Pl}$. For this plot, the quadratic variant of the symmetron model was considered, cf.~Eq.~(\ref{eq:symmetronAq}), but the same qualitative conclusions also apply to Eq.~(\ref{eq:symmetronAr}).}
\label{fig:fieldprofile14}
\end{figure}

Naive application of Eq.~(\ref{eq:alpha}) to NSs, for which $\Phi_N \sim 0.2$, would imply that already for $M_s/M_\text{Pl} \lesssim 0.1$ NSs would be screened ($\Upsilon_\text{NS} \gtrsim 10$). This expectation is confirmed for a `typical' NS. Figure \ref{fig:fieldprofile14} represents the scalar field profile inside SNSs with the same central density which, for GR, would yield a $1.4 M_\odot$ object. For $M_s \sim M_\text{Pl}$, the SNS is unscreened, with the scalar field displaying a nontrivial field gradient throughout the stellar interior. In this case, equilibrium properties such as the stellar mass and radius show $O(1)$ difference from GR. However, for $M_s \lesssim 0.1 M_\text{Pl}$, the thin shell mechanism already operates, with the fractional difference between the mass of a SNS and a GR NS dropping below 0.05\%, and similarly for other properties such as the stellar radius. 

\begin{figure*}[thb]
\centering
\includegraphics[width=0.95\textwidth]{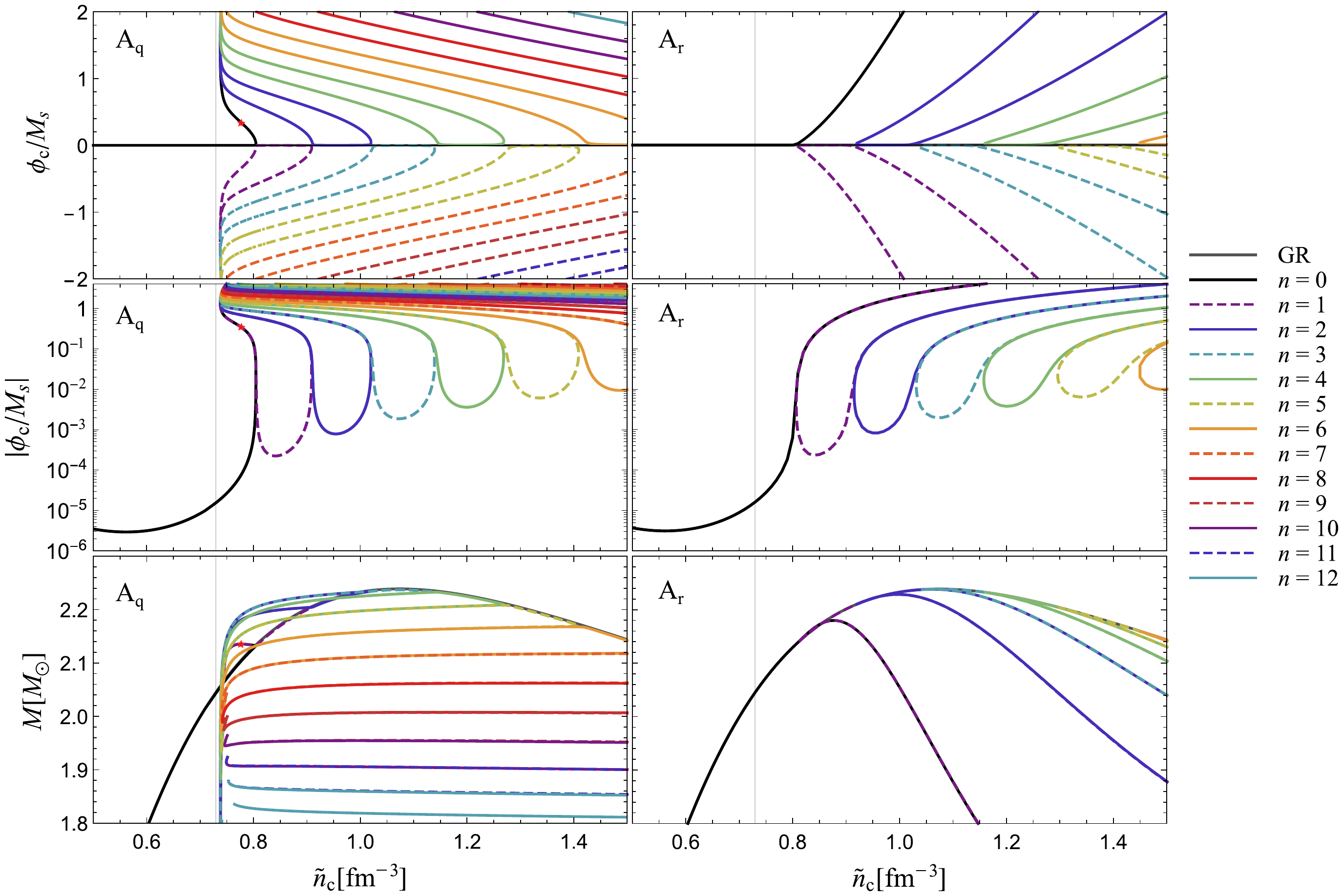}
\caption{Sequences of equilibrium solutions describing SNSs for the quadratic---left column---and regularized---right column---variants of the symmetron model, both with $M_s/M_\text{Pl} = 0.1$. The central value of the scalar field ($\phi_c$) and the total mass ($M$) are shown as functions of the central number density ($\tilde{n}_c$) in the two cases. Several branches of equilibrium solutions are identified, and classified according to the number $n$ of nodes in the scalar field profile. A vertical line is shown at $\tilde{n}_c = 0.730$ fm$^{-3}$, the critical value of the central number density above which a pressure dominated phase appears inside a GR NS. In the left column, a red star marks a marginally stable solution in the $n=0$ branch: solutions with larger values of the scalar field are found to be unstable under radial perturbations (see discussion in the main text). Note that only solutions with central value of the scalar field $|\phi_c|/M \lesssim 4$ were computed, so that curves in the bottom-left panel are incomplete, as they would contain configurations with $|\phi_c|/M > 4$.}
\label{fig:equlibrium}
\end{figure*}

However, for more massive, more compact NSs, the Newtonian reasoning above breaks down. Figure \ref{fig:equlibrium} shows sequences of equilibrium solutions for SNSs, for both the quadratic, Eq.~(\ref{eq:symmetronAq}), and regularized, Eq.~(\ref{eq:symmetronAr}), variants of the symmetron model. SNSs in both variants closely resemble those of GR as long as the trace of the energy-momentum tensor remains negative in the entire stellar interior. However, when a pressure-dominated phase appears, with $\tilde{p} > \tilde{\epsilon}/3$ and $\tilde{T} >0$ in the stellar core, SNSs become unscreened, and global properties change accordingly. 

Figure \ref{fig:equlibrium} makes clear that structural properties of unscreened SNSs may be quite sensitive to higher-order contributions to the conformal factor $A(\phi)$. In both variants, as soon as a pressure-dominated phase appears (in this case, around $\tilde{n}_c = 0.730$ fm$^{-3}$) not only the scalar field is amplified in the stellar interior, but one also finds a hierarchy of branches of equilibrium solutions, which can be classified according to the number $n$ of nodes of the scalar field profile. However, while these new branches exist {\it above} some critical central densities for the regularized variant (\ref{eq:symmetronAr}), for the quadratic one (\ref{eq:symmetronAq}) the new branches exist only \textit{below} some critical central densities. A similar behavior has also been found in the context of massless scalar-tensor theories \cite{Mendes2016,Palenzuela2016}. 

\begin{figure}[thb]
\centering
\includegraphics[width=8cm]{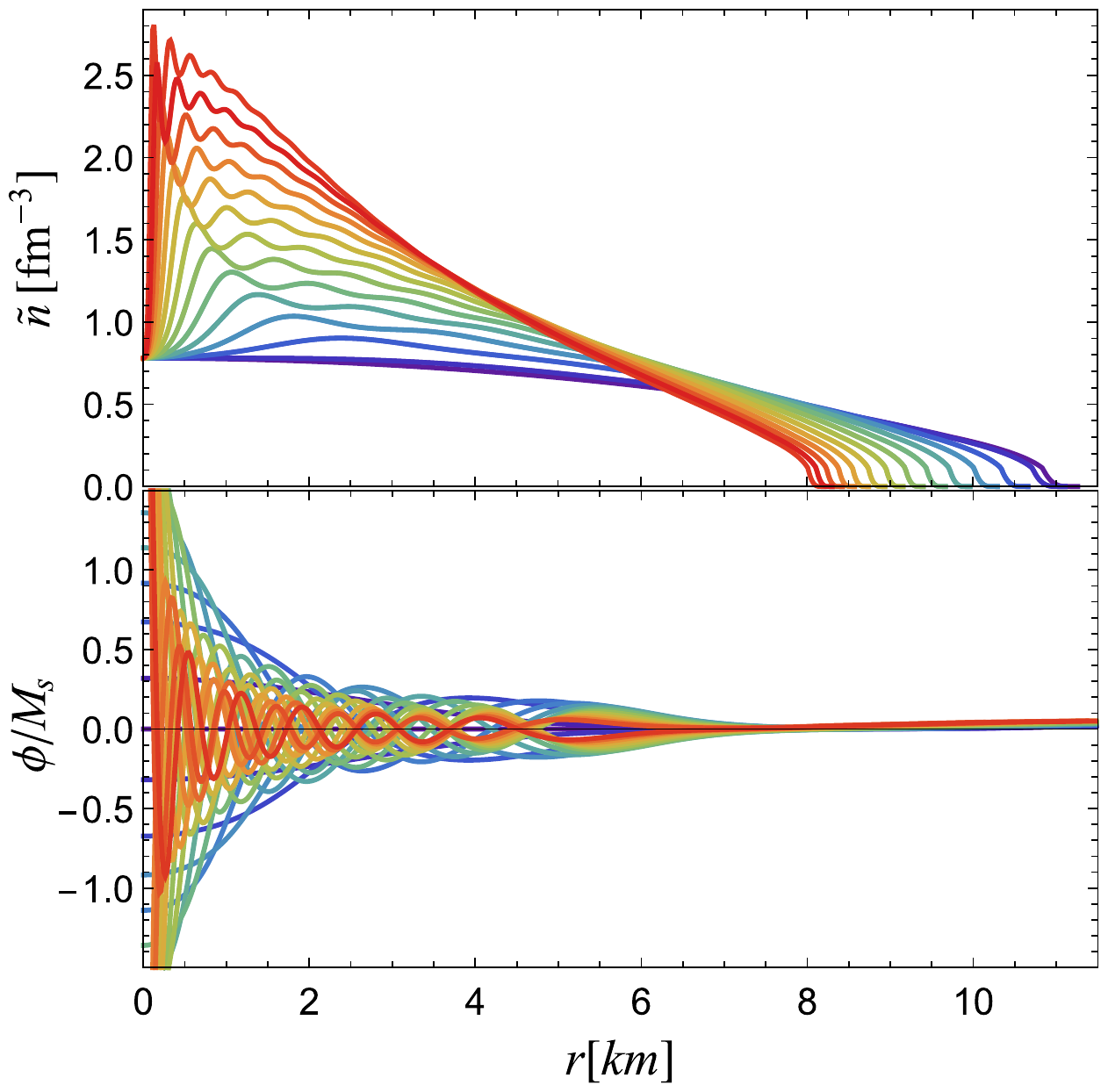}
\caption{Number density (upper panel) and scalar field (bottom panel) as functions of the radial coordinate, for SNSs with a central number density $\tilde{n}_c = 0.780$ fm${}^{-3}$, in the quadratic variant of the symmetron model with $M_s/M_\text{Pl} = 0.1$. Twenty-six solutions with $0 \leq n \leq 12$ are displayed, with the number $n$ of nodes increasing from violet to red.}
\label{fig:SNSQ}
\end{figure}

Figure \ref{fig:SNSQ} shows scalar field and density profiles for SNSs with $\tilde{n}_c = 0.780$ fm${}^{-3}$ within the quadratic variant of the model. A large, possibly infinite, number of solutions exist, but only those with $n \leq 12$ are represented in the plot. Except for the solution with $n=0$ and smallest value of $|\phi_c|$, all density profiles are non-monotonic, sustained in a delicate fluid-scalar field balance. It probably should come as no surprise that these solutions are unstable. 
Indeed, we find that all solutions with $n \neq 0$ displayed in the first column of  Fig.~\ref{fig:equlibrium} possess at least one unstable radial mode, including those with a small value of $|\phi_c|$ (the properties of which resemble those of GR NSs). For the $n=0$ branch, we find that instability sets in at $\tilde{n}_c \approx 0.776$ fm$^{-3}$, marked with a star in Fig.~\ref{fig:equlibrium}; solutions with a larger central value of the scalar field are found to be unstable.

The choice of a quadratic conformal factor (\ref{eq:symmetronAq}) is sufficiently general as long as the scalar field does not probe values close to the cutoff scale $M_s$. When this is the case, as around unscreened SNSs, one needs to care about higher order corrections to $A(\phi)$. The regularized variant in Eq.~(\ref{eq:symmetronAr}) was introduced in Ref.~\cite{Hinterbichler2011} to prevent the scalar-mediated force from becoming arbitrarily strong with increasing scalar field, by forcing it to be at most comparable to gravity. Equilibrium solutions for SNSs in this variant are displayed in the right column of Fig.~\ref{fig:equlibrium}. Again, the scalar field is strongly amplified, rising from $\sim 10^{-5} M_s$ 
for a star with the critical central density of $\tilde{n}_c = 0.730$ fm$^{-3}$ to $\sim 0.5 M_s$ at the turning point along the $n=0$ branch in the $M-\tilde{n}_c$ diagram. Furthermore, new branches of equilibrium solutions appear at higher values of the central density, characterized by an increasing number $n$ of nodes in the scalar field profile.  However, in contrast with the case of the quadratic variant (cf.~Fig.~\ref{fig:fieldprofile14}), in the regularized variant all equilibrium solutions have monotonically decreasing density profiles.

The behavior of SNSs for the regularized variant is reminiscent of the spontaneous scalarization phenomenon, a non-perturbative strong-field effect well studied since the 90s \cite{Damour1993,Salgado1998,Harada1998}. The rationale behind this effect is more easily explained for the non-interacting case with $V = 0$, and features as a main ingredient an effective coupling $\alpha(\phi)$ that is linear in $\phi$ to leading order: $\alpha(\phi) \approx \beta \phi + O(\phi^2)$. In this case, $\phi = 0$ is a solution of the scalar field equation (\ref{eq:field2}), for which Eq.~(\ref{eq:field1}) reduces to Einstein's equation; however, this solution may become unstable under a scalar field perturbation for some stellar backgrounds \cite{Harada1997,Mendes2015}. Indeed, a perturbation $\delta \phi$ around $\phi = 0$ obeys, to linear order, $\Box \delta \phi = m_\text{eff}^2 \delta \phi$, where the squared effective mass $m_\text{eff}^2 \equiv -\beta T$ was defined. If $m_\text{eff}^2$ becomes sufficiently negative, a tachyonic instability may develop. For the conformal factors (\ref{eq:symmetronAq}) and (\ref{eq:symmetronAr}), $\beta = M_s ^{-2} > 0$, and a necessary condition for the development of such tachyonic-like instability is that $T>0$; i.e., a pressure-dominated phase exists inside the star. Now, spontaneous scalarization, understood as a discontinuous change of the NS stable configuration as the baryon number of the star changes continuously \cite{Harada1998}, can be thought of as the nonlinear development of this linear tachyonic instability: As the trivial $\phi = 0$ solution becomes unstable, new equilibrium solutions develop, sustained by a nontrivial scalar field configuration. In this case, the new branches of stable equilibria appear in pairs, due to the reflection symmetry under $\phi \to -\phi$. 

In the case of SNSs, the picture described above is modified by the presence of the potential $V(\phi)$. Far away from the star, the scalar field asymptotes to the cosmological value, assumed to be the positive minimum of the effective potential, at $\phi_0 = M_s^2/M_\text{Pl}$. This breaks the degeneracy (of the $V = 0$ case) among pairs of scalarized solutions, which only differed by a $\phi \to - \phi$ transformation but otherwise had identical (macroscopic) properties. 
Now, the branch of equilibrium solutions before the onset of scalarization is smoothly connected to the $n=0$ scalarized branch, for which $\phi > 0$. At higher central densities, new branches of scalarized solutions appear in the form of connected loops (see middle row of Fig.~\ref{fig:equlibrium}). However, since local observations constrain $M_s/M_\text{Pl}$ to be small, and therefore $|\phi_0/M_s| \ll 1$, the nontrivial symmetron potential acts as a small perturbation, and pairs of scalarized solutions still exist with close macroscopic properties.
The situation has similarities to the effect of spontaneous magnetization in the presence of a small but nonzero external field, or, in a more closely related setting, to spontaneous scalarization in massless scalar-tensor theories where $\alpha_0 \equiv \alpha(0) \neq 0$ (see, e.g.,~Ref.~\cite{Rosca-Mead2020}).

\begin{figure}[thb]
\centering
\includegraphics[width=8.5cm]{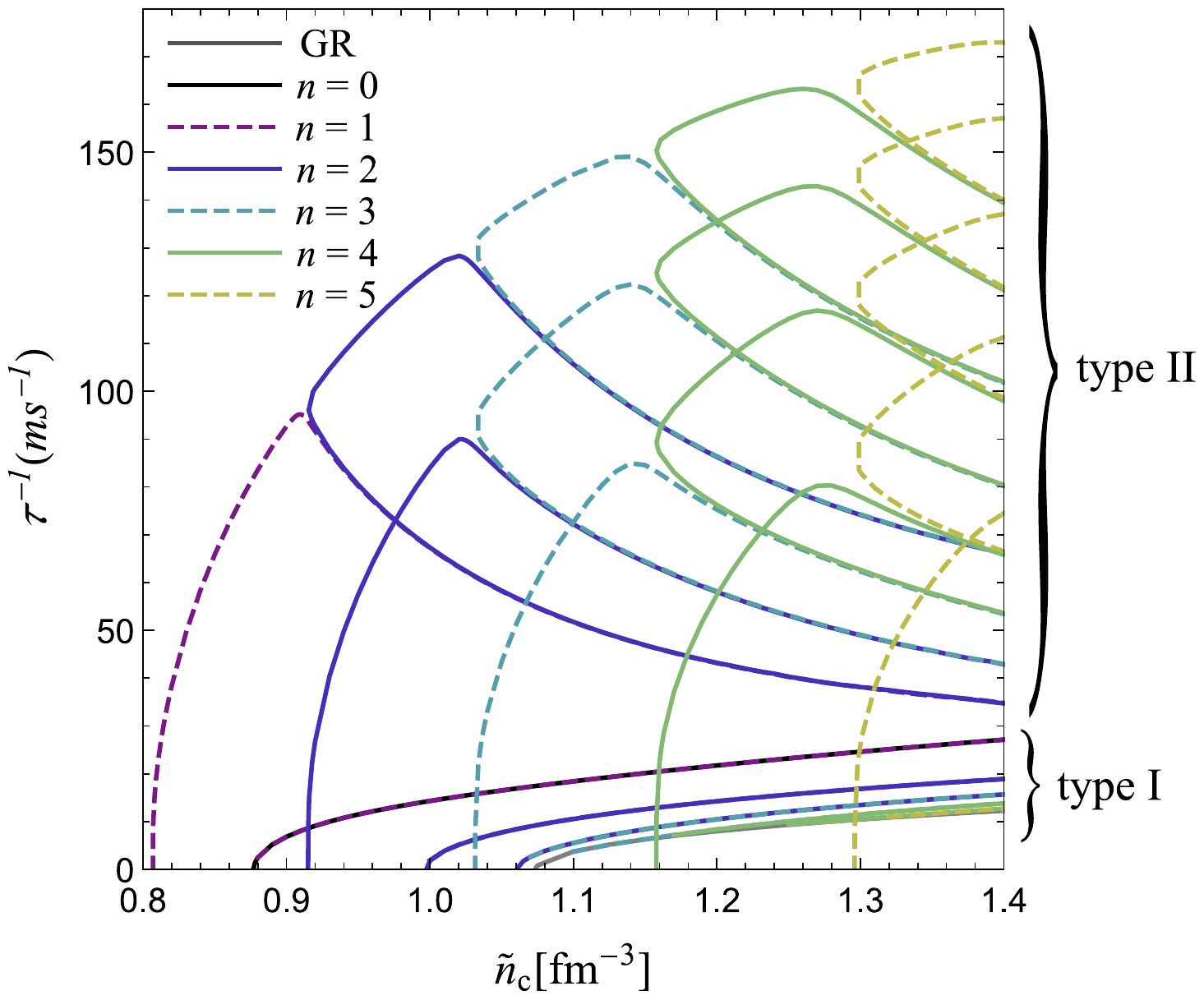}
\caption{Inverse of the instability timescale of unstable modes as a function of the central density of the corresponding equilibrium solution, for the regularized variant of the symmetron model, with $M_s/M_\text{Pl} = 0.1$. Color coding is the same as in the right column of Fig.~\ref{fig:equlibrium}, and refers to the number $n$ of nodes in the associated equilibrium solution. Modes of two types are found corresponding to a fluid-driven (type I) or a scalar-driven (type II) instability. }
\label{fig:unstableR}
\end{figure}

Additionally, we have analyzed in detail the stability of the equilibrium solutions found in the regularized variant of the symmetron model. Fig.~\ref{fig:unstableR} shows the inverse of the instability timescale $\tau^{-1} \equiv \sqrt{-\omega^2}$ for unstable modes of the equilibrium solutions displayed in the right column of Fig.~\ref{fig:equlibrium}. 
Modes of two types are found. Unstable modes of the first type (labeled ``type I'' in Fig.~\ref{fig:unstableR}) are associated with instability under gravitational collapse. For equilibrium solutions with $n = 0, 1, 2$ and for one of the legs of the $n=3$ loop, the associated unstable modes of type I emerge from a zero frequency mode at the central density corresponding to a turning point in the $M-\tilde{n}_c$ diagram; for the second leg of the $n=3$ loop and for solutions with $n > 3$, the associated unstable modes of type I emerge close to a GR unstable fundamental mode frequency. 

On the other hand, unstable modes of the second type (labeled ``type II'' in Fig.~\ref{fig:unstableR}) are associated with a scalar-field driven instability. For the $V = 0$ case, it is known that at every central density at which new scalarized solutions appear, an additional scalar mode of the trivial, $\phi = 0$, solution becomes unstable \cite{Pani2011}. Correspondingly, in the (regularized variant of the) symmetron model, we find a zero frequency mode at the critical central density at which every new loop of scalarized solutions appear. The inverse instability timescale $\tau^{-1}$ increases as one moves along (by increasing $\tilde{n}_c$) the leg with smallest value of $|\phi_c|$. When the next loop of scalarized solutions appears, it inherits the unstable mode frequencies of the previous branches (with smaller values of $n$) in the rich pattern shown in Fig.~\ref{fig:unstableR}. 
All in all, only unscreened SNSs in the $n=0$ branch and in its twin leg in the $n=1$ branch are found to be stable (up to the turning point in the $M-\tilde{n}_c$ diagram).

\begin{figure}[tb]
\centering
\includegraphics[width=8.5cm]{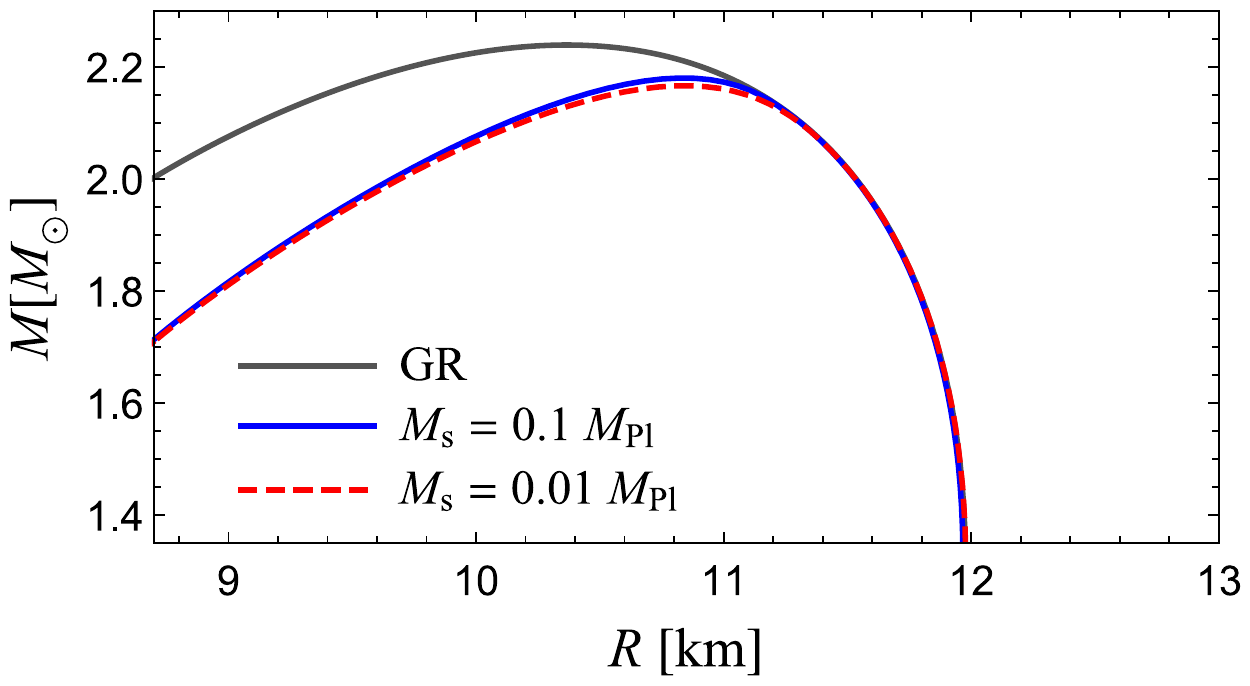}
\caption{Mass versus radius diagram for equilibrium solutions with $n=0$ in the regularized variant of the symmetron model, both for $M_s = 0.1 M_\text{Pl}$ and $M_s = 0.01 M_\text{Pl}$. The corresponding GR diagram is also represented for comparison.}
\label{fig:MvsRe0}
\end{figure}

Our discussion above has focused on the case where $M_s/M_\text{Pl} = 0.1$. Although---as argued in the beginning of this section---this value is low enough for SNSs without a pressure-dominated core to be screened, one might wonder how the picture above changes for smaller values of $M_s/M_\text{Pl}$, which are required for consistency with solar system and terrestrial observations. As $M_s/M_\text{Pl}$ decreases, $\phi_0$ is suppressed and so is the typical scalar field inside SNSs. Moreover, although the qualitative picture described above remains the same, we find that many more branches of scalarized solutions are found. For instance, while we find eight branches of equilibrium solutions for $M_s/M_\text{Pl} = 0.1$ (in the regularized variant of the symmetron model), their number rises to 63 when $M_s/M_\text{Pl} = 0.01$.
The sheer amount of equilibrium solutions, together with the increasing numerical fine-tuning needed to compute their properties, make a comprehensive analysis of their stability impracticable for $M_s/M_\text{Pl} \lesssim 0.01$. However, several of our findings should extrapolate for that case. For instance, Fig.~\ref{fig:MvsRe0} shows a mass-radius diagram for the $n=0$ branch of equilibrium solutions in the regularized variant of the symmetron model, both for $M_s = 0.1 M_\text{Pl}$ and $M_s = 0.01 M_\text{Pl}$, where we can see that their properties are only weakly dependent on $M_s$.

\section{Conclusions} \label{sec:conclusion}

A partial failure of typical screening mechanisms may occur in the cores of the most massive, most compact neutron stars observed in Nature if the nuclear EOS is such that a pressure-dominated phase occurs in their interior \cite{Aguiar2020,Dima2021}. Here we have examined equilibrium and stability properties of such unscreened NSs (described by the ENG EOS) in two variants of the symmetron model, characterized by the conformal factors (\ref{eq:symmetronAq}) or (\ref{eq:symmetronAr}), for which the effective coupling $\alpha(\phi)$ between the symmetron and matter differs in its higher than linear behavior. 

In both cases, NSs are screened and have nearly identical properties to their GR counterparts before the critical density for the appearance of a pressure-dominated phase. However, their properties differ much more strongly once this critical density is reached, as shown in Fig.~\ref{fig:equlibrium}. 
For the regularized variant, the stable branches of equilibrium solutions (with $n=0$ and one leg with $n=1$) have a lower maximum mass than that of GR, a decrease of $\sim 4.6 \%$, similar to that found for the chameleon and environmentally dependent dilaton models in Ref.~\cite{Aguiar2020}. For the quadratic variant, stable equilibrium solutions cease to exist soon after the critical density for the appearance of a pressure-dominated phase, promoting an abrupt cut before the maximum mass is reached. In both cases, the spectrum of unstable modes presents a rich structure, as shown in Fig.~\ref{fig:unstableR} for the regularized variant. 

That NSs with pressure-dominated cores should exist in Nature is an intriguing possibility. Interestingly, the existence of a strong correlation between $p_c/\epsilon_c$ (the ratio between pressure and energy density at the NS core) and macroscopic properties such as the NS compactness or tidal deformability \cite{Saes2021,Podkowka2018} makes it possible to access the existence of a pressure-dominated phase ($p_c/\epsilon_c > 1/3$) by measuring properties of the most massive NSs. In fact, there is a nonnegligible probability that the heavy pulsar J0030+0451 displays such a phase \cite{Saes2021}. If more accurate measurements of massive NSs support this conclusion, these systems could be interesting probes of scalar-tensor theories with screening mechanisms, as presently shown for the symmetron model.

\acknowledgments
BA acknowledges financial support from the Coordination of Superior Level Staff Improvement (CAPES); RM and FTF acknowledge financial support from the National Council for Scientific and Technological Development (CNPq), and RM also from the Carlos Chagas Filho Research Support Foundation (FAPERJ).

\bibliography{library}

\begin{thebibliography}{60}%
\makeatletter
\providecommand \@ifxundefined [1]{%
 \@ifx{#1\undefined}
}%
\providecommand \@ifnum [1]{%
 \ifnum #1\expandafter \@firstoftwo
 \else \expandafter \@secondoftwo
 \fi
}%
\providecommand \@ifx [1]{%
 \ifx #1\expandafter \@firstoftwo
 \else \expandafter \@secondoftwo
 \fi
}%
\providecommand \natexlab [1]{#1}%
\providecommand \enquote  [1]{``#1''}%
\providecommand \bibnamefont  [1]{#1}%
\providecommand \bibfnamefont [1]{#1}%
\providecommand \citenamefont [1]{#1}%
\providecommand \href@noop [0]{\@secondoftwo}%
\providecommand \href [0]{\begingroup \@sanitize@url \@href}%
\providecommand \@href[1]{\@@startlink{#1}\@@href}%
\providecommand \@@href[1]{\endgroup#1\@@endlink}%
\providecommand \@sanitize@url [0]{\catcode `\\12\catcode `\$12\catcode
  `\&12\catcode `\#12\catcode `\^12\catcode `\_12\catcode `\%12\relax}%
\providecommand \@@startlink[1]{}%
\providecommand \@@endlink[0]{}%
\providecommand \url  [0]{\begingroup\@sanitize@url \@url }%
\providecommand \@url [1]{\endgroup\@href {#1}{\urlprefix }}%
\providecommand \urlprefix  [0]{URL }%
\providecommand \Eprint [0]{\href }%
\providecommand \doibase [0]{https://doi.org/}%
\providecommand \selectlanguage [0]{\@gobble}%
\providecommand \bibinfo  [0]{\@secondoftwo}%
\providecommand \bibfield  [0]{\@secondoftwo}%
\providecommand \translation [1]{[#1]}%
\providecommand \BibitemOpen [0]{}%
\providecommand \bibitemStop [0]{}%
\providecommand \bibitemNoStop [0]{.\EOS\space}%
\providecommand \EOS [0]{\spacefactor3000\relax}%
\providecommand \BibitemShut  [1]{\csname bibitem#1\endcsname}%
\let\auto@bib@innerbib\@empty
\bibitem [{\citenamefont {{\"{O}}zel}\ and\ \citenamefont
  {Freire}(2016)}]{Ozel2016}%
  \BibitemOpen
  \bibfield  {author} {\bibinfo {author} {\bibfnamefont {F.}~\bibnamefont
  {{\"{O}}zel}}\ and\ \bibinfo {author} {\bibfnamefont {P.}~\bibnamefont
  {Freire}},\ }\bibfield  {title} {\bibinfo {title} {{Masses, Radii, and the
  Equation of State of Neutron Stars}},\ }\href@noop {} {\bibfield  {journal}
  {\bibinfo  {journal} {Annu. Rev. Astron. Astrophys.}\ }\textbf {\bibinfo
  {volume} {54}},\ \bibinfo {pages} {401} (\bibinfo {year} {2016})}\BibitemShut
  {NoStop}%
\bibitem [{\citenamefont {Abbott}\ \emph {et~al.}(2019)\citenamefont {Abbott}
  \emph {et~al.}}]{TheLIGOScientificCollaboration2018}%
  \BibitemOpen
  \bibfield  {author} {\bibinfo {author} {\bibfnamefont {B.~P.}\ \bibnamefont
  {Abbott}} \emph {et~al.},\ }\bibfield  {title} {\bibinfo {title} {{GWTC-1: A
  Gravitational-Wave Transient Catalog of Compact Binary Mergers Observed by
  LIGO and Virgo during the First and Second Observing Runs}},\ }\href
  {http://arxiv.org/abs/1811.12907
  https://link.aps.org/doi/10.1103/PhysRevX.9.031040} {\bibfield  {journal}
  {\bibinfo  {journal} {Phys. Rev. X}\ }\textbf {\bibinfo {volume} {9}},\
  \bibinfo {pages} {031040} (\bibinfo {year} {2019})}\BibitemShut {NoStop}%
\bibitem [{\citenamefont {Abbott}\ \emph
  {et~al.}(2021{\natexlab{a}})\citenamefont {Abbott}, \citenamefont {Abbott},
  \citenamefont {Abraham} \emph {et~al.}}]{Abbott2021}%
  \BibitemOpen
  \bibfield  {author} {\bibinfo {author} {\bibfnamefont {R.}~\bibnamefont
  {Abbott}}, \bibinfo {author} {\bibfnamefont {T.~D.}\ \bibnamefont {Abbott}},
  \bibinfo {author} {\bibnamefont {Abraham}}, \emph {et~al.},\ }\bibfield
  {title} {\bibinfo {title} {{GWTC-2: Compact Binary Coalescences Observed by
  LIGO and Virgo during the First Half of the Third Observing Run}},\ }\href
  {https://journals.aps.org/prx/abstract/10.1103/PhysRevX.11.021053} {\bibfield
   {journal} {\bibinfo  {journal} {Physical Review X}\ }\textbf {\bibinfo
  {volume} {11}},\ \bibinfo {pages} {021053} (\bibinfo {year}
  {2021}{\natexlab{a}})}\BibitemShut {NoStop}%
\bibitem [{\citenamefont {Abbott}\ \emph
  {et~al.}(2021{\natexlab{b}})\citenamefont {Abbott} \emph
  {et~al.}}]{theligoscientificcollaboration2021gwtc3}%
  \BibitemOpen
  \bibfield  {author} {\bibinfo {author} {\bibfnamefont {R.}~\bibnamefont
  {Abbott}} \emph {et~al.},\ }\bibfield  {title} {\bibinfo {title} {{GWTC-3:
  Compact Binary Coalescences Observed by LIGO and Virgo During the Second Part
  of the Third Observing Run}},\ }\href@noop {} {\bibfield  {journal} {\bibinfo
   {journal} {arXiv e-prints}\ ,\ \bibinfo {pages} {arXiv:2111.03606}}
  (\bibinfo {year} {2021}{\natexlab{b}})}\BibitemShut {NoStop}%
\bibitem [{\citenamefont {{Pietroni}}(2005)}]{Pietroni2005PRD}%
  \BibitemOpen
  \bibfield  {author} {\bibinfo {author} {\bibfnamefont {M.}~\bibnamefont
  {{Pietroni}}},\ }\bibfield  {title} {\bibinfo {title} {{Dark energy
  condensation}},\ }\href@noop {} {\bibfield  {journal} {\bibinfo  {journal}
  {Physical Review D}\ }\textbf {\bibinfo {volume} {72}},\ \bibinfo {eid}
  {043535} (\bibinfo {year} {2005})}\BibitemShut {NoStop}%
\bibitem [{\citenamefont {{Olive}}\ and\ \citenamefont
  {{Pospelov}}(2008)}]{Olive2008PRD}%
  \BibitemOpen
  \bibfield  {author} {\bibinfo {author} {\bibfnamefont {K.~A.}\ \bibnamefont
  {{Olive}}}\ and\ \bibinfo {author} {\bibfnamefont {M.}~\bibnamefont
  {{Pospelov}}},\ }\bibfield  {title} {\bibinfo {title} {{Environmental
  dependence of masses and coupling constants}},\ }\href@noop {} {\bibfield
  {journal} {\bibinfo  {journal} {Physical Review D}\ }\textbf {\bibinfo
  {volume} {77}},\ \bibinfo {eid} {043524} (\bibinfo {year}
  {2008})}\BibitemShut {NoStop}%
\bibitem [{\citenamefont {{Khoury}}(2010)}]{Khoury2010}%
  \BibitemOpen
  \bibfield  {author} {\bibinfo {author} {\bibfnamefont {J.}~\bibnamefont
  {{Khoury}}},\ }\bibfield  {title} {\bibinfo {title} {{Theories of Dark Energy
  with Screening Mechanisms}},\ }\href@noop {} {\bibfield  {journal} {\bibinfo
  {journal} {arXiv e-prints}\ ,\ \bibinfo {eid} {arXiv:1011.5909}} (\bibinfo
  {year} {2010})}\BibitemShut {NoStop}%
\bibitem [{\citenamefont {Damour}\ and\ \citenamefont
  {Esposito-Far{\`{e}}se}(1992)}]{Damour1992}%
  \BibitemOpen
  \bibfield  {author} {\bibinfo {author} {\bibfnamefont {T.}~\bibnamefont
  {Damour}}\ and\ \bibinfo {author} {\bibfnamefont {G.}~\bibnamefont
  {Esposito-Far{\`{e}}se}},\ }\bibfield  {title} {\bibinfo {title}
  {Tensor-multi-scalar theories of gravitation},\ }\href
  {http://iopscience.iop.org/0264-9381/9/9/015} {\bibfield  {journal} {\bibinfo
   {journal} {Class. Quantum Gravity}\ }\textbf {\bibinfo {volume} {9}},\
  \bibinfo {pages} {2093} (\bibinfo {year} {1992})}\BibitemShut {NoStop}%
\bibitem [{\citenamefont {{Damour}}\ and\ \citenamefont
  {{Polyakov}}(1994)}]{Damour1994NuPB}%
  \BibitemOpen
  \bibfield  {author} {\bibinfo {author} {\bibfnamefont {T.}~\bibnamefont
  {{Damour}}}\ and\ \bibinfo {author} {\bibfnamefont {A.~M.}\ \bibnamefont
  {{Polyakov}}},\ }\bibfield  {title} {\bibinfo {title} {{The string dilation
  and a least coupling principle}},\ }\href@noop {} {\bibfield  {journal}
  {\bibinfo  {journal} {Nuclear Physics B}\ }\textbf {\bibinfo {volume}
  {423}},\ \bibinfo {pages} {532} (\bibinfo {year} {1994})}\BibitemShut
  {NoStop}%
\bibitem [{\citenamefont {{Baker}}\ \emph {et~al.}(2021)\citenamefont
  {{Baker}}, \citenamefont {{Barreira}}, \citenamefont {{Desmond}},
  \citenamefont {{Ferreira}}, \citenamefont {{Jain}}, \citenamefont {{Koyama}},
  \citenamefont {{Li}}, \citenamefont {{Lombriser}}, \citenamefont {{Nicola}},
  \citenamefont {{Sakstein}},\ and\ \citenamefont {{Schmidt}}}]{baker2019}%
  \BibitemOpen
  \bibfield  {author} {\bibinfo {author} {\bibfnamefont {T.}~\bibnamefont
  {{Baker}}}, \bibinfo {author} {\bibfnamefont {A.}~\bibnamefont {{Barreira}}},
  \bibinfo {author} {\bibfnamefont {H.}~\bibnamefont {{Desmond}}}, \bibinfo
  {author} {\bibfnamefont {P.}~\bibnamefont {{Ferreira}}}, \bibinfo {author}
  {\bibfnamefont {B.}~\bibnamefont {{Jain}}}, \bibinfo {author} {\bibfnamefont
  {K.}~\bibnamefont {{Koyama}}}, \bibinfo {author} {\bibfnamefont
  {B.}~\bibnamefont {{Li}}}, \bibinfo {author} {\bibfnamefont {L.}~\bibnamefont
  {{Lombriser}}}, \bibinfo {author} {\bibfnamefont {A.}~\bibnamefont
  {{Nicola}}}, \bibinfo {author} {\bibfnamefont {J.}~\bibnamefont
  {{Sakstein}}},\ and\ \bibinfo {author} {\bibfnamefont {F.}~\bibnamefont
  {{Schmidt}}},\ }\bibfield  {title} {\bibinfo {title} {{Novel Probes Project:
  Tests of gravity on astrophysical scales}},\ }\href@noop {} {\bibfield
  {journal} {\bibinfo  {journal} {Reviews of Modern Physics}\ }\textbf
  {\bibinfo {volume} {93}},\ \bibinfo {eid} {015003} (\bibinfo {year}
  {2021})}\BibitemShut {NoStop}%
\bibitem [{\citenamefont {{Koyama}}(2016)}]{Koyama2016}%
  \BibitemOpen
  \bibfield  {author} {\bibinfo {author} {\bibfnamefont {K.}~\bibnamefont
  {{Koyama}}},\ }\bibfield  {title} {\bibinfo {title} {{Cosmological tests of
  modified gravity}},\ }\href@noop {} {\bibfield  {journal} {\bibinfo
  {journal} {Reports on Progress in Physics}\ }\textbf {\bibinfo {volume}
  {79}},\ \bibinfo {eid} {046902} (\bibinfo {year} {2016})}\BibitemShut
  {NoStop}%
\bibitem [{\citenamefont {{Bull}}\ \emph {et~al.}(2016)\citenamefont {{Bull}},
  \citenamefont {{Akrami}}, \citenamefont {{Adamek}}, \citenamefont {{Baker}},
  \citenamefont {{Bellini}}, \citenamefont {{Beltr{\'a}n Jim{\'e}nez}},
  \citenamefont {{Bentivegna}}, \citenamefont {{Camera}}, \citenamefont
  {{Clesse}}, \citenamefont {{Davis}}, \citenamefont {{Di Dio}}, \citenamefont
  {{Enander}}, \citenamefont {{Heavens}}, \citenamefont {{Heisenberg}},
  \citenamefont {{Hu}}, \citenamefont {{Llinares}}, \citenamefont {{Maartens}},
  \citenamefont {{M{\"o}rtsell}}, \citenamefont {{Nadathur}}, \citenamefont
  {{Noller}}, \citenamefont {{Pasechnik}}, \citenamefont {{Pawlowski}},
  \citenamefont {{Pereira}}, \citenamefont {{Quartin}}, \citenamefont
  {{Ricciardone}}, \citenamefont {{Riemer-S{\o}rensen}}, \citenamefont
  {{Rinaldi}}, \citenamefont {{Sakstein}}, \citenamefont {{Saltas}},
  \citenamefont {{Salzano}}, \citenamefont {{Sawicki}}, \citenamefont
  {{Solomon}}, \citenamefont {{Spolyar}}, \citenamefont {{Starkman}},
  \citenamefont {{Steer}}, \citenamefont {{Tereno}}, \citenamefont {{Verde}},
  \citenamefont {{Villaescusa-Navarro}}, \citenamefont {{von Strauss}},\ and\
  \citenamefont {{Winther}}}]{Bull2016}%
  \BibitemOpen
  \bibfield  {author} {\bibinfo {author} {\bibfnamefont {P.}~\bibnamefont
  {{Bull}}}, \bibinfo {author} {\bibfnamefont {Y.}~\bibnamefont {{Akrami}}},
  \bibinfo {author} {\bibfnamefont {J.}~\bibnamefont {{Adamek}}}, \bibinfo
  {author} {\bibfnamefont {T.}~\bibnamefont {{Baker}}}, \bibinfo {author}
  {\bibfnamefont {E.}~\bibnamefont {{Bellini}}}, \bibinfo {author}
  {\bibfnamefont {J.}~\bibnamefont {{Beltr{\'a}n Jim{\'e}nez}}}, \bibinfo
  {author} {\bibfnamefont {E.}~\bibnamefont {{Bentivegna}}}, \bibinfo {author}
  {\bibfnamefont {S.}~\bibnamefont {{Camera}}}, \bibinfo {author}
  {\bibfnamefont {S.}~\bibnamefont {{Clesse}}}, \bibinfo {author}
  {\bibfnamefont {J.~H.}\ \bibnamefont {{Davis}}}, \bibinfo {author}
  {\bibfnamefont {E.}~\bibnamefont {{Di Dio}}}, \bibinfo {author}
  {\bibfnamefont {J.}~\bibnamefont {{Enander}}}, \bibinfo {author}
  {\bibfnamefont {A.}~\bibnamefont {{Heavens}}}, \bibinfo {author}
  {\bibfnamefont {L.}~\bibnamefont {{Heisenberg}}}, \bibinfo {author}
  {\bibfnamefont {B.}~\bibnamefont {{Hu}}}, \bibinfo {author} {\bibfnamefont
  {C.}~\bibnamefont {{Llinares}}}, \bibinfo {author} {\bibfnamefont
  {R.}~\bibnamefont {{Maartens}}}, \bibinfo {author} {\bibfnamefont
  {E.}~\bibnamefont {{M{\"o}rtsell}}}, \bibinfo {author} {\bibfnamefont
  {S.}~\bibnamefont {{Nadathur}}}, \bibinfo {author} {\bibfnamefont
  {J.}~\bibnamefont {{Noller}}}, \bibinfo {author} {\bibfnamefont
  {R.}~\bibnamefont {{Pasechnik}}}, \bibinfo {author} {\bibfnamefont {M.~S.}\
  \bibnamefont {{Pawlowski}}}, \bibinfo {author} {\bibfnamefont {T.~S.}\
  \bibnamefont {{Pereira}}}, \bibinfo {author} {\bibfnamefont {M.}~\bibnamefont
  {{Quartin}}}, \bibinfo {author} {\bibfnamefont {A.}~\bibnamefont
  {{Ricciardone}}}, \bibinfo {author} {\bibfnamefont {S.}~\bibnamefont
  {{Riemer-S{\o}rensen}}}, \bibinfo {author} {\bibfnamefont {M.}~\bibnamefont
  {{Rinaldi}}}, \bibinfo {author} {\bibfnamefont {J.}~\bibnamefont
  {{Sakstein}}}, \bibinfo {author} {\bibfnamefont {I.~D.}\ \bibnamefont
  {{Saltas}}}, \bibinfo {author} {\bibfnamefont {V.}~\bibnamefont {{Salzano}}},
  \bibinfo {author} {\bibfnamefont {I.}~\bibnamefont {{Sawicki}}}, \bibinfo
  {author} {\bibfnamefont {A.~R.}\ \bibnamefont {{Solomon}}}, \bibinfo {author}
  {\bibfnamefont {D.}~\bibnamefont {{Spolyar}}}, \bibinfo {author}
  {\bibfnamefont {G.~D.}\ \bibnamefont {{Starkman}}}, \bibinfo {author}
  {\bibfnamefont {D.}~\bibnamefont {{Steer}}}, \bibinfo {author} {\bibfnamefont
  {I.}~\bibnamefont {{Tereno}}}, \bibinfo {author} {\bibfnamefont
  {L.}~\bibnamefont {{Verde}}}, \bibinfo {author} {\bibfnamefont
  {F.}~\bibnamefont {{Villaescusa-Navarro}}}, \bibinfo {author} {\bibfnamefont
  {M.}~\bibnamefont {{von Strauss}}},\ and\ \bibinfo {author} {\bibfnamefont
  {H.~A.}\ \bibnamefont {{Winther}}},\ }\bibfield  {title} {\bibinfo {title}
  {{Beyond {\ensuremath{\Lambda}} CDM: Problems, solutions, and the road
  ahead}},\ }\href@noop {} {\bibfield  {journal} {\bibinfo  {journal} {Physics
  of the Dark Universe}\ }\textbf {\bibinfo {volume} {12}},\ \bibinfo {pages}
  {56} (\bibinfo {year} {2016})}\BibitemShut {NoStop}%
\bibitem [{\citenamefont {Clifton}\ \emph {et~al.}(2012)\citenamefont
  {Clifton}, \citenamefont {Ferreira}, \citenamefont {Padilla},\ and\
  \citenamefont {Skordis}}]{Clifton2012}%
  \BibitemOpen
  \bibfield  {author} {\bibinfo {author} {\bibfnamefont {T.}~\bibnamefont
  {Clifton}}, \bibinfo {author} {\bibfnamefont {P.~G.}\ \bibnamefont
  {Ferreira}}, \bibinfo {author} {\bibfnamefont {A.}~\bibnamefont {Padilla}},\
  and\ \bibinfo {author} {\bibfnamefont {C.}~\bibnamefont {Skordis}},\
  }\bibfield  {title} {\bibinfo {title} {{Modified gravity and cosmology}},\
  }\href {http://www.sciencedirect.com/science/article/pii/S0370157312000105}
  {\bibfield  {journal} {\bibinfo  {journal} {Phys. Rep.}\ }\textbf {\bibinfo
  {volume} {513}},\ \bibinfo {pages} {1} (\bibinfo {year} {2012})}\BibitemShut
  {NoStop}%
\bibitem [{\citenamefont {{Joyce}}\ \emph {et~al.}(2015)\citenamefont
  {{Joyce}}, \citenamefont {{Jain}}, \citenamefont {{Khoury}},\ and\
  \citenamefont {{Trodden}}}]{Joyce2015}%
  \BibitemOpen
  \bibfield  {author} {\bibinfo {author} {\bibfnamefont {A.}~\bibnamefont
  {{Joyce}}}, \bibinfo {author} {\bibfnamefont {B.}~\bibnamefont {{Jain}}},
  \bibinfo {author} {\bibfnamefont {J.}~\bibnamefont {{Khoury}}},\ and\
  \bibinfo {author} {\bibfnamefont {M.}~\bibnamefont {{Trodden}}},\ }\bibfield
  {title} {\bibinfo {title} {{Beyond the cosmological standard model}},\
  }\href@noop {} {\bibfield  {journal} {\bibinfo  {journal} {Physics Report}\
  }\textbf {\bibinfo {volume} {568}},\ \bibinfo {pages} {1} (\bibinfo {year}
  {2015})}\BibitemShut {NoStop}%
\bibitem [{\citenamefont {Will}(2006)}]{Will2006}%
  \BibitemOpen
  \bibfield  {author} {\bibinfo {author} {\bibfnamefont {C.~M.}\ \bibnamefont
  {Will}},\ }\bibfield  {title} {\bibinfo {title} {{The confrontation between
  general relativity and experiment}},\ }\href@noop {} {\bibfield  {journal}
  {\bibinfo  {journal} {Living Rev. Relativ.}\ }\textbf {\bibinfo {volume}
  {9}},\ \bibinfo {pages} {1} (\bibinfo {year} {2006})}\BibitemShut {NoStop}%
\bibitem [{\citenamefont {Sakstein}(2018)}]{Sakstein2020}%
  \BibitemOpen
  \bibfield  {author} {\bibinfo {author} {\bibfnamefont {J.}~\bibnamefont
  {Sakstein}},\ }\bibfield  {title} {\bibinfo {title} {{Astrophysical tests of
  screened modified gravity}},\ }\href@noop {} {\bibfield  {journal} {\bibinfo
  {journal} {Int. J. Mod. Phys. D}\ }\textbf {\bibinfo {volume} {27}},\
  \bibinfo {pages} {1848008} (\bibinfo {year} {2018})}\BibitemShut {NoStop}%
\bibitem [{\citenamefont {Khoury}\ and\ \citenamefont
  {Weltman}(2004{\natexlab{a}})}]{Khoury2004}%
  \BibitemOpen
  \bibfield  {author} {\bibinfo {author} {\bibfnamefont {J.}~\bibnamefont
  {Khoury}}\ and\ \bibinfo {author} {\bibfnamefont {A.}~\bibnamefont
  {Weltman}},\ }\bibfield  {title} {\bibinfo {title} {{Chameleon cosmology}},\
  }\href {https://link.aps.org/doi/10.1103/PhysRevD.69.044026} {\bibfield
  {journal} {\bibinfo  {journal} {Phys. Rev. D}\ }\textbf {\bibinfo {volume}
  {69}},\ \bibinfo {pages} {044026} (\bibinfo {year}
  {2004}{\natexlab{a}})}\BibitemShut {NoStop}%
\bibitem [{\citenamefont {Khoury}\ and\ \citenamefont
  {Weltman}(2004{\natexlab{b}})}]{Khoury2004a}%
  \BibitemOpen
  \bibfield  {author} {\bibinfo {author} {\bibfnamefont {J.}~\bibnamefont
  {Khoury}}\ and\ \bibinfo {author} {\bibfnamefont {A.}~\bibnamefont
  {Weltman}},\ }\bibfield  {title} {\bibinfo {title} {{Chameleon Fields:
  Awaiting Surprises for Tests of Gravity in Space}},\ }\href
  {https://link.aps.org/doi/10.1103/PhysRevLett.93.171104} {\bibfield
  {journal} {\bibinfo  {journal} {Phys. Rev. Lett.}\ }\textbf {\bibinfo
  {volume} {93}},\ \bibinfo {pages} {171104} (\bibinfo {year}
  {2004}{\natexlab{b}})}\BibitemShut {NoStop}%
\bibitem [{\citenamefont {Hinterbichler}\ and\ \citenamefont
  {Khoury}(2010)}]{Hinterbichler2010}%
  \BibitemOpen
  \bibfield  {author} {\bibinfo {author} {\bibfnamefont {K.}~\bibnamefont
  {Hinterbichler}}\ and\ \bibinfo {author} {\bibfnamefont {J.}~\bibnamefont
  {Khoury}},\ }\bibfield  {title} {\bibinfo {title} {{Screening Long-Range
  Forces through Local Symmetry Restoration}},\ }\href
  {http://link.aps.org/doi/10.1103/PhysRevLett.104.231301} {\bibfield
  {journal} {\bibinfo  {journal} {Phys. Rev. Lett.}\ }\textbf {\bibinfo
  {volume} {104}},\ \bibinfo {pages} {231301} (\bibinfo {year}
  {2010})}\BibitemShut {NoStop}%
\bibitem [{\citenamefont {Hinterbichler}\ \emph {et~al.}(2011)\citenamefont
  {Hinterbichler}, \citenamefont {Khoury}, \citenamefont {Levy},\ and\
  \citenamefont {Matas}}]{Hinterbichler2011}%
  \BibitemOpen
  \bibfield  {author} {\bibinfo {author} {\bibfnamefont {K.}~\bibnamefont
  {Hinterbichler}}, \bibinfo {author} {\bibfnamefont {J.}~\bibnamefont
  {Khoury}}, \bibinfo {author} {\bibfnamefont {A.}~\bibnamefont {Levy}},\ and\
  \bibinfo {author} {\bibfnamefont {A.}~\bibnamefont {Matas}},\ }\bibfield
  {title} {\bibinfo {title} {Symmetron cosmology},\ }\href
  {https://link.aps.org/doi/10.1103/PhysRevD.84.103521} {\bibfield  {journal}
  {\bibinfo  {journal} {Phys. Rev. D}\ }\textbf {\bibinfo {volume} {84}},\
  \bibinfo {pages} {103521} (\bibinfo {year} {2011})}\BibitemShut {NoStop}%
\bibitem [{\citenamefont {Brax}\ \emph
  {et~al.}(2012{\natexlab{a}})\citenamefont {Brax}, \citenamefont {Davis},
  \citenamefont {Li},\ and\ \citenamefont {Winther}}]{Brax2012}%
  \BibitemOpen
  \bibfield  {author} {\bibinfo {author} {\bibfnamefont {P.}~\bibnamefont
  {Brax}}, \bibinfo {author} {\bibfnamefont {A.-C.}\ \bibnamefont {Davis}},
  \bibinfo {author} {\bibfnamefont {B.}~\bibnamefont {Li}},\ and\ \bibinfo
  {author} {\bibfnamefont {H.~A.}\ \bibnamefont {Winther}},\ }\bibfield
  {title} {\bibinfo {title} {{Unified description of screened modified
  gravity}},\ }\href {https://link.aps.org/doi/10.1103/PhysRevD.86.044015}
  {\bibfield  {journal} {\bibinfo  {journal} {Phys. Rev. D}\ }\textbf {\bibinfo
  {volume} {86}},\ \bibinfo {pages} {044015} (\bibinfo {year}
  {2012}{\natexlab{a}})}\BibitemShut {NoStop}%
\bibitem [{\citenamefont {Brax}\ \emph
  {et~al.}(2012{\natexlab{b}})\citenamefont {Brax}, \citenamefont {Davis},\
  and\ \citenamefont {Li}}]{Brax201238}%
  \BibitemOpen
  \bibfield  {author} {\bibinfo {author} {\bibfnamefont {P.}~\bibnamefont
  {Brax}}, \bibinfo {author} {\bibfnamefont {A.-C.}\ \bibnamefont {Davis}},\
  and\ \bibinfo {author} {\bibfnamefont {B.}~\bibnamefont {Li}},\ }\bibfield
  {title} {\bibinfo {title} {Modified gravity tomography},\ }\href
  {https://www.sciencedirect.com/science/article/pii/S0370269312008374}
  {\bibfield  {journal} {\bibinfo  {journal} {Physics Letters B}\ }\textbf
  {\bibinfo {volume} {715}},\ \bibinfo {pages} {38} (\bibinfo {year}
  {2012}{\natexlab{b}})}\BibitemShut {NoStop}%
\bibitem [{\citenamefont {{Burrage}}\ \emph {et~al.}(2016)\citenamefont
  {{Burrage}}, \citenamefont {{Copeland}},\ and\ \citenamefont
  {{Millington}}}]{Burrage2016PRL}%
  \BibitemOpen
  \bibfield  {author} {\bibinfo {author} {\bibfnamefont {C.}~\bibnamefont
  {{Burrage}}}, \bibinfo {author} {\bibfnamefont {E.~J.}\ \bibnamefont
  {{Copeland}}},\ and\ \bibinfo {author} {\bibfnamefont {P.}~\bibnamefont
  {{Millington}}},\ }\bibfield  {title} {\bibinfo {title} {{Radiative Screening
  of Fifth Forces}},\ }\href@noop {} {\bibfield  {journal} {\bibinfo  {journal}
  {Physical Review Letters}\ }\textbf {\bibinfo {volume} {117}},\ \bibinfo
  {eid} {211102} (\bibinfo {year} {2016})}\BibitemShut {NoStop}%
\bibitem [{\citenamefont {Brax}\ \emph {et~al.}(2011)\citenamefont {Brax},
  \citenamefont {van~de Bruck}, \citenamefont {Davis}, \citenamefont {Li},
  \citenamefont {Schmauch},\ and\ \citenamefont {Shaw}}]{Brax2011}%
  \BibitemOpen
  \bibfield  {author} {\bibinfo {author} {\bibfnamefont {P.}~\bibnamefont
  {Brax}}, \bibinfo {author} {\bibfnamefont {C.}~\bibnamefont {van~de Bruck}},
  \bibinfo {author} {\bibfnamefont {A.-C.}\ \bibnamefont {Davis}}, \bibinfo
  {author} {\bibfnamefont {B.}~\bibnamefont {Li}}, \bibinfo {author}
  {\bibfnamefont {B.}~\bibnamefont {Schmauch}},\ and\ \bibinfo {author}
  {\bibfnamefont {D.~J.}\ \bibnamefont {Shaw}},\ }\bibfield  {title} {\bibinfo
  {title} {Linear growth of structure in the symmetron model},\ }\href
  {https://link.aps.org/doi/10.1103/PhysRevD.84.123524} {\bibfield  {journal}
  {\bibinfo  {journal} {Phys. Rev. D}\ }\textbf {\bibinfo {volume} {84}},\
  \bibinfo {pages} {123524} (\bibinfo {year} {2011})}\BibitemShut {NoStop}%
\bibitem [{\citenamefont {{Brax}}\ \emph {et~al.}(2012)\citenamefont {{Brax}},
  \citenamefont {{Davis}}, \citenamefont {{Li}}, \citenamefont {{Winther}},\
  and\ \citenamefont {{Zhao}}}]{Brax2012JCAP}%
  \BibitemOpen
  \bibfield  {author} {\bibinfo {author} {\bibfnamefont {P.}~\bibnamefont
  {{Brax}}}, \bibinfo {author} {\bibfnamefont {A.-C.}\ \bibnamefont {{Davis}}},
  \bibinfo {author} {\bibfnamefont {B.}~\bibnamefont {{Li}}}, \bibinfo {author}
  {\bibfnamefont {H.~A.}\ \bibnamefont {{Winther}}},\ and\ \bibinfo {author}
  {\bibfnamefont {G.-B.}\ \bibnamefont {{Zhao}}},\ }\bibfield  {title}
  {\bibinfo {title} {{Systematic simulations of modified gravity: symmetron and
  dilaton models}},\ }\href@noop {} {\bibfield  {journal} {\bibinfo  {journal}
  {JCAP}\ }\textbf {\bibinfo {volume} {2012}}\bibinfo  {number} { (10)},\
  \bibinfo {eid} {002}}\BibitemShut {NoStop}%
\bibitem [{\citenamefont {Davis}\ \emph {et~al.}(2012)\citenamefont {Davis},
  \citenamefont {Li}, \citenamefont {Mota},\ and\ \citenamefont
  {Winther}}]{Davis2012}%
  \BibitemOpen
\bibfield  {number} {  }\bibfield  {author} {\bibinfo {author} {\bibfnamefont
  {A.-C.}\ \bibnamefont {Davis}}, \bibinfo {author} {\bibfnamefont
  {B.}~\bibnamefont {Li}}, \bibinfo {author} {\bibfnamefont {D.~F.}\
  \bibnamefont {Mota}},\ and\ \bibinfo {author} {\bibfnamefont {H.~A.}\
  \bibnamefont {Winther}},\ }\bibfield  {title} {\bibinfo {title} {Structure
  formation in the symmetron model},\ }\href
  {https://doi.org/10.1088/0004-637x/748/1/61} {\bibfield  {journal} {\bibinfo
  {journal} {The Astrophysical Journal}\ }\textbf {\bibinfo {volume} {748}},\
  \bibinfo {pages} {61} (\bibinfo {year} {2012})}\BibitemShut {NoStop}%
\bibitem [{\citenamefont {Llinares}\ and\ \citenamefont
  {Mota}(2013)}]{Llinares2013}%
  \BibitemOpen
  \bibfield  {author} {\bibinfo {author} {\bibfnamefont {C.}~\bibnamefont
  {Llinares}}\ and\ \bibinfo {author} {\bibfnamefont {D.~F.}\ \bibnamefont
  {Mota}},\ }\bibfield  {title} {\bibinfo {title} {Releasing scalar fields:
  Cosmological simulations of scalar-tensor theories for gravity beyond the
  static approximation},\ }\href
  {https://link.aps.org/doi/10.1103/PhysRevLett.110.161101} {\bibfield
  {journal} {\bibinfo  {journal} {Phys. Rev. Lett.}\ }\textbf {\bibinfo
  {volume} {110}},\ \bibinfo {pages} {161101} (\bibinfo {year}
  {2013})}\BibitemShut {NoStop}%
\bibitem [{\citenamefont {Llinares}\ and\ \citenamefont
  {Mota}(2014)}]{Llinares2014}%
  \BibitemOpen
  \bibfield  {author} {\bibinfo {author} {\bibfnamefont {C.}~\bibnamefont
  {Llinares}}\ and\ \bibinfo {author} {\bibfnamefont {D.~F.}\ \bibnamefont
  {Mota}},\ }\bibfield  {title} {\bibinfo {title} {{Cosmological simulations of
  screened modified gravity out of the static approximation: effects on matter
  distribution}},\ }\href@noop {} {\bibfield  {journal} {\bibinfo  {journal}
  {Phys. Rev. D}\ }\textbf {\bibinfo {volume} {89}},\ \bibinfo {pages} {084023}
  (\bibinfo {year} {2014})}\BibitemShut {NoStop}%
\bibitem [{\citenamefont {{Mota}}(2018)}]{Mota2018IJMPD}%
  \BibitemOpen
  \bibfield  {author} {\bibinfo {author} {\bibfnamefont {D.~F.}\ \bibnamefont
  {{Mota}}},\ }\bibfield  {title} {\bibinfo {title} {{Probing screened modified
  gravity with nonlinear structure formation}},\ }\href@noop {} {\bibfield
  {journal} {\bibinfo  {journal} {International Journal of Modern Physics D}\
  }\textbf {\bibinfo {volume} {27}},\ \bibinfo {eid} {1830003} (\bibinfo {year}
  {2018})}\BibitemShut {NoStop}%
\bibitem [{\citenamefont {Clampitt}\ \emph {et~al.}(2012)\citenamefont
  {Clampitt}, \citenamefont {Jain},\ and\ \citenamefont
  {Khoury}}]{Clampitt2012}%
  \BibitemOpen
  \bibfield  {author} {\bibinfo {author} {\bibfnamefont {J.}~\bibnamefont
  {Clampitt}}, \bibinfo {author} {\bibfnamefont {B.}~\bibnamefont {Jain}},\
  and\ \bibinfo {author} {\bibfnamefont {J.}~\bibnamefont {Khoury}},\
  }\bibfield  {title} {\bibinfo {title} {Halo scale predictions of symmetron
  modified gravity},\ }\href {https://doi.org/10.1088/1475-7516/2012/01/030}
  {\bibfield  {journal} {\bibinfo  {journal} {Journal of Cosmology and
  Astroparticle Physics}\ }\textbf {\bibinfo {volume} {2012}}\bibinfo  {number}
  { (01)},\ \bibinfo {pages} {030}}\BibitemShut {NoStop}%
\bibitem [{\citenamefont {{Taddei}}\ \emph {et~al.}(2014)\citenamefont
  {{Taddei}}, \citenamefont {{Catena}},\ and\ \citenamefont
  {{Pietroni}}}]{Taddei2014PRD}%
  \BibitemOpen
\bibfield  {number} {  }\bibfield  {author} {\bibinfo {author} {\bibfnamefont
  {L.}~\bibnamefont {{Taddei}}}, \bibinfo {author} {\bibfnamefont
  {R.}~\bibnamefont {{Catena}}},\ and\ \bibinfo {author} {\bibfnamefont
  {M.}~\bibnamefont {{Pietroni}}},\ }\bibfield  {title} {\bibinfo {title}
  {{Spherical collapse and halo mass function in the symmetron model}},\ }\href
  {https://doi.org/10.1103/PhysRevD.89.023523} {\bibfield  {journal} {\bibinfo
  {journal} {Physical Review D}\ }\textbf {\bibinfo {volume} {89}},\ \bibinfo
  {eid} {023523} (\bibinfo {year} {2014})}\BibitemShut {NoStop}%
\bibitem [{\citenamefont {{Contigiani}}\ \emph {et~al.}(2019)\citenamefont
  {{Contigiani}}, \citenamefont {{Vardanyan}},\ and\ \citenamefont
  {{Silvestri}}}]{Contigiani2019PRD}%
  \BibitemOpen
  \bibfield  {author} {\bibinfo {author} {\bibfnamefont {O.}~\bibnamefont
  {{Contigiani}}}, \bibinfo {author} {\bibfnamefont {V.}~\bibnamefont
  {{Vardanyan}}},\ and\ \bibinfo {author} {\bibfnamefont {A.}~\bibnamefont
  {{Silvestri}}},\ }\bibfield  {title} {\bibinfo {title} {{Splashback radius in
  symmetron gravity}},\ }\href {https://doi.org/10.1103/PhysRevD.99.064030}
  {\bibfield  {journal} {\bibinfo  {journal} {Physical Review D}\ }\textbf
  {\bibinfo {volume} {99}},\ \bibinfo {eid} {064030} (\bibinfo {year}
  {2019})}\BibitemShut {NoStop}%
\bibitem [{\citenamefont {Burrage}\ \emph {et~al.}(2017)\citenamefont
  {Burrage}, \citenamefont {Copeland},\ and\ \citenamefont
  {Millington}}]{Burrage2017}%
  \BibitemOpen
  \bibfield  {author} {\bibinfo {author} {\bibfnamefont {C.}~\bibnamefont
  {Burrage}}, \bibinfo {author} {\bibfnamefont {E.~J.}\ \bibnamefont
  {Copeland}},\ and\ \bibinfo {author} {\bibfnamefont {P.}~\bibnamefont
  {Millington}},\ }\bibfield  {title} {\bibinfo {title} {Radial acceleration
  relation from symmetron fifth forces},\ }\href
  {https://link.aps.org/doi/10.1103/PhysRevD.95.064050} {\bibfield  {journal}
  {\bibinfo  {journal} {Phys. Rev. D}\ }\textbf {\bibinfo {volume} {95}},\
  \bibinfo {pages} {064050} (\bibinfo {year} {2017})}\BibitemShut {NoStop}%
\bibitem [{\citenamefont {{O'Hare}}\ and\ \citenamefont
  {{Burrage}}(2018)}]{Ohare2018PRD}%
  \BibitemOpen
  \bibfield  {author} {\bibinfo {author} {\bibfnamefont {C.~A.~J.}\
  \bibnamefont {{O'Hare}}}\ and\ \bibinfo {author} {\bibfnamefont
  {C.}~\bibnamefont {{Burrage}}},\ }\bibfield  {title} {\bibinfo {title}
  {{Stellar kinematics from the symmetron fifth force in the Milky Way disk}},\
  }\href {https://doi.org/10.1103/PhysRevD.98.064019} {\bibfield  {journal}
  {\bibinfo  {journal} {Physical Review D}\ }\textbf {\bibinfo {volume} {98}},\
  \bibinfo {eid} {064019} (\bibinfo {year} {2018})}\BibitemShut {NoStop}%
\bibitem [{\citenamefont {Burrage}\ \emph {et~al.}(2019)\citenamefont
  {Burrage}, \citenamefont {Copeland}, \citenamefont {K\"ading},\ and\
  \citenamefont {Millington}}]{Burrage2019}%
  \BibitemOpen
  \bibfield  {author} {\bibinfo {author} {\bibfnamefont {C.}~\bibnamefont
  {Burrage}}, \bibinfo {author} {\bibfnamefont {E.~J.}\ \bibnamefont
  {Copeland}}, \bibinfo {author} {\bibfnamefont {C.}~\bibnamefont {K\"ading}},\
  and\ \bibinfo {author} {\bibfnamefont {P.}~\bibnamefont {Millington}},\
  }\bibfield  {title} {\bibinfo {title} {Symmetron scalar fields: Modified
  gravity, dark matter, or both?},\ }\href
  {https://link.aps.org/doi/10.1103/PhysRevD.99.043539} {\bibfield  {journal}
  {\bibinfo  {journal} {Phys. Rev. D}\ }\textbf {\bibinfo {volume} {99}},\
  \bibinfo {pages} {043539} (\bibinfo {year} {2019})}\BibitemShut {NoStop}%
\bibitem [{\citenamefont {Upadhye}(2013)}]{Upadhye2013}%
  \BibitemOpen
  \bibfield  {author} {\bibinfo {author} {\bibfnamefont {A.}~\bibnamefont
  {Upadhye}},\ }\bibfield  {title} {\bibinfo {title} {Symmetron dark energy in
  laboratory experiments},\ }\href
  {https://link.aps.org/doi/10.1103/PhysRevLett.110.031301} {\bibfield
  {journal} {\bibinfo  {journal} {Phys. Rev. Lett.}\ }\textbf {\bibinfo
  {volume} {110}},\ \bibinfo {pages} {031301} (\bibinfo {year}
  {2013})}\BibitemShut {NoStop}%
\bibitem [{\citenamefont {Burrage}\ \emph {et~al.}(2016)\citenamefont
  {Burrage}, \citenamefont {Kuribayashi-Coleman}, \citenamefont {Stevenson},\
  and\ \citenamefont {Thrussell}}]{Burrage2016b}%
  \BibitemOpen
  \bibfield  {author} {\bibinfo {author} {\bibfnamefont {C.}~\bibnamefont
  {Burrage}}, \bibinfo {author} {\bibfnamefont {A.}~\bibnamefont
  {Kuribayashi-Coleman}}, \bibinfo {author} {\bibfnamefont {J.}~\bibnamefont
  {Stevenson}},\ and\ \bibinfo {author} {\bibfnamefont {B.}~\bibnamefont
  {Thrussell}},\ }\bibfield  {title} {\bibinfo {title} {Constraining symmetron
  fields with atom interferometry},\ }\href
  {https://doi.org/10.1088/1475-7516/2016/12/041} {\bibfield  {journal}
  {\bibinfo  {journal} {JCAP}\ }\textbf {\bibinfo {volume} {2016}}\bibinfo
  {number} { (12)},\ \bibinfo {pages} {041}}\BibitemShut {NoStop}%
\bibitem [{\citenamefont {{Cronenberg}}\ \emph {et~al.}(2018)\citenamefont
  {{Cronenberg}}, \citenamefont {{Brax}}, \citenamefont {{Filter}},
  \citenamefont {{Geltenbort}}, \citenamefont {{Jenke}}, \citenamefont
  {{Pignol}}, \citenamefont {{Pitschmann}}, \citenamefont {{Thalhammer}},\ and\
  \citenamefont {{Abele}}}]{Cronenberg2018NatPh}%
  \BibitemOpen
\bibfield  {number} {  }\bibfield  {author} {\bibinfo {author} {\bibfnamefont
  {G.}~\bibnamefont {{Cronenberg}}}, \bibinfo {author} {\bibfnamefont
  {P.}~\bibnamefont {{Brax}}}, \bibinfo {author} {\bibfnamefont
  {H.}~\bibnamefont {{Filter}}}, \bibinfo {author} {\bibfnamefont
  {P.}~\bibnamefont {{Geltenbort}}}, \bibinfo {author} {\bibfnamefont
  {T.}~\bibnamefont {{Jenke}}}, \bibinfo {author} {\bibfnamefont
  {G.}~\bibnamefont {{Pignol}}}, \bibinfo {author} {\bibfnamefont
  {M.}~\bibnamefont {{Pitschmann}}}, \bibinfo {author} {\bibfnamefont
  {M.}~\bibnamefont {{Thalhammer}}},\ and\ \bibinfo {author} {\bibfnamefont
  {H.}~\bibnamefont {{Abele}}},\ }\bibfield  {title} {\bibinfo {title}
  {{Acoustic Rabi oscillations between gravitational quantum states and impact
  on symmetron dark energy}},\ }\href
  {https://doi.org/10.1038/s41567-018-0205-x} {\bibfield  {journal} {\bibinfo
  {journal} {Nature Physics}\ }\textbf {\bibinfo {volume} {14}},\ \bibinfo
  {pages} {1022} (\bibinfo {year} {2018})}\BibitemShut {NoStop}%
\bibitem [{\citenamefont {{Chiow}}\ and\ \citenamefont
  {{Yu}}(2020)}]{Chiow2020PRvD}%
  \BibitemOpen
  \bibfield  {author} {\bibinfo {author} {\bibfnamefont {S.-w.}\ \bibnamefont
  {{Chiow}}}\ and\ \bibinfo {author} {\bibfnamefont {N.}~\bibnamefont {{Yu}}},\
  }\bibfield  {title} {\bibinfo {title} {{Constraining symmetron dark energy
  using atom interferometry}},\ }\href
  {https://doi.org/10.1103/PhysRevD.101.083501} {\bibfield  {journal} {\bibinfo
   {journal} {Physical Review D}\ }\textbf {\bibinfo {volume} {101}},\ \bibinfo
  {eid} {083501} (\bibinfo {year} {2020})}\BibitemShut {NoStop}%
\bibitem [{\citenamefont {Elder}\ \emph {et~al.}(2020)\citenamefont {Elder},
  \citenamefont {Vardanyan}, \citenamefont {Akrami}, \citenamefont {Brax},
  \citenamefont {Davis},\ and\ \citenamefont {Decca}}]{Elder2020}%
  \BibitemOpen
  \bibfield  {author} {\bibinfo {author} {\bibfnamefont {B.}~\bibnamefont
  {Elder}}, \bibinfo {author} {\bibfnamefont {V.}~\bibnamefont {Vardanyan}},
  \bibinfo {author} {\bibfnamefont {Y.}~\bibnamefont {Akrami}}, \bibinfo
  {author} {\bibfnamefont {P.}~\bibnamefont {Brax}}, \bibinfo {author}
  {\bibfnamefont {A.-C.}\ \bibnamefont {Davis}},\ and\ \bibinfo {author}
  {\bibfnamefont {R.~S.}\ \bibnamefont {Decca}},\ }\bibfield  {title} {\bibinfo
  {title} {Classical symmetron force in casimir experiments},\ }\href
  {https://link.aps.org/doi/10.1103/PhysRevD.101.064065} {\bibfield  {journal}
  {\bibinfo  {journal} {Phys. Rev. D}\ }\textbf {\bibinfo {volume} {101}},\
  \bibinfo {pages} {064065} (\bibinfo {year} {2020})}\BibitemShut {NoStop}%
\bibitem [{\citenamefont {Burrage}\ and\ \citenamefont
  {Sakstein}(2018)}]{Burrage2018}%
  \BibitemOpen
  \bibfield  {author} {\bibinfo {author} {\bibfnamefont {C.}~\bibnamefont
  {Burrage}}\ and\ \bibinfo {author} {\bibfnamefont {J.}~\bibnamefont
  {Sakstein}},\ }\bibfield  {title} {\bibinfo {title} {{Tests of chameleon
  gravity}},\ }\href {http://link.springer.com/10.1007/s41114-018-0011-x}
  {\bibfield  {journal} {\bibinfo  {journal} {Living Rev. Relativ.}\ }\textbf
  {\bibinfo {volume} {21}},\ \bibinfo {pages} {1} (\bibinfo {year}
  {2018})}\BibitemShut {NoStop}%
\bibitem [{\citenamefont {Babichev}\ and\ \citenamefont
  {Langlois}(2010)}]{Babichev2010}%
  \BibitemOpen
  \bibfield  {author} {\bibinfo {author} {\bibfnamefont {E.}~\bibnamefont
  {Babichev}}\ and\ \bibinfo {author} {\bibfnamefont {D.}~\bibnamefont
  {Langlois}},\ }\bibfield  {title} {\bibinfo {title} {{Relativistic stars in
  f(R) and scalar-tensor theories}},\ }\href
  {https://link.aps.org/doi/10.1103/PhysRevD.81.124051} {\bibfield  {journal}
  {\bibinfo  {journal} {Phys. Rev. D}\ }\textbf {\bibinfo {volume} {81}},\
  \bibinfo {pages} {124051} (\bibinfo {year} {2010})}\BibitemShut {NoStop}%
\bibitem [{\citenamefont {Brax}\ \emph {et~al.}(2017)\citenamefont {Brax},
  \citenamefont {Davis},\ and\ \citenamefont {Jha}}]{Brax2017}%
  \BibitemOpen
  \bibfield  {author} {\bibinfo {author} {\bibfnamefont {P.}~\bibnamefont
  {Brax}}, \bibinfo {author} {\bibfnamefont {A.-C.}\ \bibnamefont {Davis}},\
  and\ \bibinfo {author} {\bibfnamefont {R.}~\bibnamefont {Jha}},\ }\bibfield
  {title} {\bibinfo {title} {{Neutron stars in screened modified gravity:
  Chameleon versus dilaton}},\ }\href
  {http://link.aps.org/doi/10.1103/PhysRevD.95.083514} {\bibfield  {journal}
  {\bibinfo  {journal} {Phys. Rev. D}\ }\textbf {\bibinfo {volume} {95}},\
  \bibinfo {pages} {083514} (\bibinfo {year} {2017})}\BibitemShut {NoStop}%
\bibitem [{\citenamefont {de~Aguiar}\ and\ \citenamefont
  {Mendes}(2020)}]{Aguiar2020}%
  \BibitemOpen
  \bibfield  {author} {\bibinfo {author} {\bibfnamefont {B.~F.}\ \bibnamefont
  {de~Aguiar}}\ and\ \bibinfo {author} {\bibfnamefont {R.~F.~P.}\ \bibnamefont
  {Mendes}},\ }\bibfield  {title} {\bibinfo {title} {Highly compact neutron
  stars and screening mechanisms: Equilibrium and stability},\ }\href
  {https://link.aps.org/doi/10.1103/PhysRevD.102.024064} {\bibfield  {journal}
  {\bibinfo  {journal} {Phys. Rev. D}\ }\textbf {\bibinfo {volume} {102}},\
  \bibinfo {pages} {024064} (\bibinfo {year} {2020})}\BibitemShut {NoStop}%
\bibitem [{\citenamefont {Podkowka}\ \emph {et~al.}(2018)\citenamefont
  {Podkowka}, \citenamefont {Mendes},\ and\ \citenamefont
  {Poisson}}]{Podkowka2018}%
  \BibitemOpen
  \bibfield  {author} {\bibinfo {author} {\bibfnamefont {D.~M.}\ \bibnamefont
  {Podkowka}}, \bibinfo {author} {\bibfnamefont {R.~F.~P.}\ \bibnamefont
  {Mendes}},\ and\ \bibinfo {author} {\bibfnamefont {E.}~\bibnamefont
  {Poisson}},\ }\bibfield  {title} {\bibinfo {title} {{Trace of the
  energy-momentum tensor and macroscopic properties of neutron stars}},\ }\href
  {https://link.aps.org/doi/10.1103/PhysRevD.98.064057} {\bibfield  {journal}
  {\bibinfo  {journal} {Phys. Rev. D}\ }\textbf {\bibinfo {volume} {98}},\
  \bibinfo {pages} {064057} (\bibinfo {year} {2018})}\BibitemShut {NoStop}%
\bibitem [{\citenamefont {Saes}\ and\ \citenamefont {Mendes}(2021)}]{Saes2021}%
  \BibitemOpen
  \bibfield  {author} {\bibinfo {author} {\bibfnamefont {J.}~\bibnamefont
  {Saes}}\ and\ \bibinfo {author} {\bibfnamefont {R.~F.~P.}\ \bibnamefont
  {Mendes}},\ }\bibfield  {title} {\bibinfo {title} {An
  equation-of-state-insensitive measure of neutron star stiffness},\
  }\href@noop {} {\bibfield  {journal} {\bibinfo  {journal} {arXiv e-prints}\
  ,\ \bibinfo {pages} {2109.11571}} (\bibinfo {year} {2021})}\BibitemShut
  {NoStop}%
\bibitem [{\citenamefont {Dima}\ \emph {et~al.}(2021)\citenamefont {Dima},
  \citenamefont {Bezares},\ and\ \citenamefont {Barausse}}]{Dima2021}%
  \BibitemOpen
  \bibfield  {author} {\bibinfo {author} {\bibfnamefont {A.}~\bibnamefont
  {Dima}}, \bibinfo {author} {\bibfnamefont {M.}~\bibnamefont {Bezares}},\ and\
  \bibinfo {author} {\bibfnamefont {E.}~\bibnamefont {Barausse}},\ }\bibfield
  {title} {\bibinfo {title} {Dynamical chameleon neutron stars: Stability,
  radial oscillations, and scalar radiation in spherical symmetry},\ }\href
  {https://link.aps.org/doi/10.1103/PhysRevD.104.084017} {\bibfield  {journal}
  {\bibinfo  {journal} {Phys. Rev. D}\ }\textbf {\bibinfo {volume} {104}},\
  \bibinfo {pages} {084017} (\bibinfo {year} {2021})}\BibitemShut {NoStop}%
\bibitem [{\citenamefont {Jain}\ \emph {et~al.}(2013)\citenamefont {Jain},
  \citenamefont {Vikram},\ and\ \citenamefont {Sakstein}}]{Jain2013}%
  \BibitemOpen
  \bibfield  {author} {\bibinfo {author} {\bibfnamefont {B.}~\bibnamefont
  {Jain}}, \bibinfo {author} {\bibfnamefont {V.}~\bibnamefont {Vikram}},\ and\
  \bibinfo {author} {\bibfnamefont {J.}~\bibnamefont {Sakstein}},\ }\bibfield
  {title} {\bibinfo {title} {Astrophysical tests of modified gravity:
  constraints from distance indicators in the nearby universe},\ }\href
  {https://iopscience.iop.org/article/10.1088/0004-637X/779/1/39} {\bibfield
  {journal} {\bibinfo  {journal} {Astrophys. J.}\ }\textbf {\bibinfo {volume}
  {779}},\ \bibinfo {pages} {39} (\bibinfo {year} {2013})}\BibitemShut
  {NoStop}%
\bibitem [{\citenamefont {Engvik}\ \emph {et~al.}(1996)\citenamefont {Engvik},
  \citenamefont {Osnes}, \citenamefont {Hjorth-Jensen}, \citenamefont {Bao},\
  and\ \citenamefont {Ostgaard}}]{Engvik1995}%
  \BibitemOpen
  \bibfield  {author} {\bibinfo {author} {\bibfnamefont {L.}~\bibnamefont
  {Engvik}}, \bibinfo {author} {\bibfnamefont {E.}~\bibnamefont {Osnes}},
  \bibinfo {author} {\bibfnamefont {M.}~\bibnamefont {Hjorth-Jensen}}, \bibinfo
  {author} {\bibfnamefont {G.}~\bibnamefont {Bao}},\ and\ \bibinfo {author}
  {\bibfnamefont {E.}~\bibnamefont {Ostgaard}},\ }\bibfield  {title} {\bibinfo
  {title} {{Asymmetric Nuclear Matter and Neutron Star Properties}},\ }\href
  {http://arxiv.org/abs/nucl-th/9509016 http://dx.doi.org/10.1086/177827
  http://adsabs.harvard.edu/doi/10.1086/177827} {\bibfield  {journal} {\bibinfo
   {journal} {Astrophys. J.}\ }\textbf {\bibinfo {volume} {469}},\ \bibinfo
  {pages} {794} (\bibinfo {year} {1996})}\BibitemShut {NoStop}%
\bibitem [{\citenamefont {Read}\ \emph {et~al.}(2009)\citenamefont {Read},
  \citenamefont {Lackey}, \citenamefont {Owen},\ and\ \citenamefont
  {Friedman}}]{Read2009}%
  \BibitemOpen
  \bibfield  {author} {\bibinfo {author} {\bibfnamefont {J.}~\bibnamefont
  {Read}}, \bibinfo {author} {\bibfnamefont {B.}~\bibnamefont {Lackey}},
  \bibinfo {author} {\bibfnamefont {B.}~\bibnamefont {Owen}},\ and\ \bibinfo
  {author} {\bibfnamefont {J.~L.}\ \bibnamefont {Friedman}},\ }\bibfield
  {title} {\bibinfo {title} {{Constraints on a phenomenologically parametrized
  neutron-star equation of state}},\ }\href
  {https://doi.org/10.1103/PhysRevD.79.124032} {\bibfield  {journal} {\bibinfo
  {journal} {Phys. Rev. D}\ }\textbf {\bibinfo {volume} {79}},\ \bibinfo
  {pages} {124032} (\bibinfo {year} {2009})}\BibitemShut {NoStop}%
\bibitem [{Note1()}]{Note1}%
  \BibitemOpen
  \bibinfo {note} {Explicit expressions for ${\protect \bf M}$ and ${\protect
  \bf F}$ are available at \protect \url
  {https://bitbucket.org/raissafpmendes/symmetron_neutron_stars.git}.}\BibitemShut
  {Stop}%
\bibitem [{\citenamefont {Mendes}\ and\ \citenamefont
  {Ortiz}(2016)}]{Mendes2016}%
  \BibitemOpen
  \bibfield  {author} {\bibinfo {author} {\bibfnamefont {R.~F.~P.}\
  \bibnamefont {Mendes}}\ and\ \bibinfo {author} {\bibfnamefont
  {N.}~\bibnamefont {Ortiz}},\ }\bibfield  {title} {\bibinfo {title} {{Highly
  compact neutron stars in scalar-tensor theories of gravity: Spontaneous
  scalarization versus gravitational collapse}},\ }\href
  {http://link.aps.org/doi/10.1103/PhysRevD.93.124035} {\bibfield  {journal}
  {\bibinfo  {journal} {Phys. Rev. D}\ }\textbf {\bibinfo {volume} {93}},\
  \bibinfo {pages} {124035} (\bibinfo {year} {2016})}\BibitemShut {NoStop}%
\bibitem [{\citenamefont {Palenzuela}\ and\ \citenamefont
  {Liebling}(2016)}]{Palenzuela2016}%
  \BibitemOpen
  \bibfield  {author} {\bibinfo {author} {\bibfnamefont {C.}~\bibnamefont
  {Palenzuela}}\ and\ \bibinfo {author} {\bibfnamefont {S.~L.}\ \bibnamefont
  {Liebling}},\ }\bibfield  {title} {\bibinfo {title} {{Constraining
  scalar-tensor theories of gravity from the most massive neutron stars}},\
  }\href {http://journals.aps.org/prd/abstract/10.1103/PhysRevD.93.044009}
  {\bibfield  {journal} {\bibinfo  {journal} {Phys. Rev. D}\ }\textbf {\bibinfo
  {volume} {93}},\ \bibinfo {pages} {044009} (\bibinfo {year}
  {2016})}\BibitemShut {NoStop}%
\bibitem [{\citenamefont {Damour}\ and\ \citenamefont
  {Esposito-Far{\`{e}}se}(1993)}]{Damour1993}%
  \BibitemOpen
  \bibfield  {author} {\bibinfo {author} {\bibfnamefont {T.}~\bibnamefont
  {Damour}}\ and\ \bibinfo {author} {\bibfnamefont {G.}~\bibnamefont
  {Esposito-Far{\`{e}}se}},\ }\bibfield  {title} {\bibinfo {title}
  {{Nonperturbative strong-field effects in tensor-scalar theories of
  gravitation}},\ }\href {http://link.aps.org/doi/10.1103/PhysRevLett.70.2220}
  {\bibfield  {journal} {\bibinfo  {journal} {Phys. Rev. Lett.}\ }\textbf
  {\bibinfo {volume} {70}},\ \bibinfo {pages} {2220} (\bibinfo {year}
  {1993})}\BibitemShut {NoStop}%
\bibitem [{\citenamefont {Salgado}\ \emph {et~al.}(1998)\citenamefont
  {Salgado}, \citenamefont {Sudarsky},\ and\ \citenamefont
  {Nucamendi}}]{Salgado1998}%
  \BibitemOpen
  \bibfield  {author} {\bibinfo {author} {\bibfnamefont {M.}~\bibnamefont
  {Salgado}}, \bibinfo {author} {\bibfnamefont {D.}~\bibnamefont {Sudarsky}},\
  and\ \bibinfo {author} {\bibfnamefont {U.}~\bibnamefont {Nucamendi}},\
  }\bibfield  {title} {\bibinfo {title} {{Spontaneous scalarization}},\ }\href
  {http://link.aps.org/doi/10.1103/PhysRevD.58.124003} {\bibfield  {journal}
  {\bibinfo  {journal} {Phys. Rev. D}\ }\textbf {\bibinfo {volume} {58}},\
  \bibinfo {pages} {124003} (\bibinfo {year} {1998})}\BibitemShut {NoStop}%
\bibitem [{\citenamefont {Harada}(1998)}]{Harada1998}%
  \BibitemOpen
  \bibfield  {author} {\bibinfo {author} {\bibfnamefont {T.}~\bibnamefont
  {Harada}},\ }\bibfield  {title} {\bibinfo {title} {{Neutron stars in
  scalar-tensor theories of gravity and catastrophe theory}},\ }\href
  {http://link.aps.org/doi/10.1103/PhysRevD.57.4802} {\bibfield  {journal}
  {\bibinfo  {journal} {Phys. Rev. D}\ }\textbf {\bibinfo {volume} {57}},\
  \bibinfo {pages} {4802} (\bibinfo {year} {1998})}\BibitemShut {NoStop}%
\bibitem [{\citenamefont {Harada}(1997)}]{Harada1997}%
  \BibitemOpen
  \bibfield  {author} {\bibinfo {author} {\bibfnamefont {T.}~\bibnamefont
  {Harada}},\ }\bibfield  {title} {\bibinfo {title} {{Stability Analysis of
  Spherically Symmetric Star in Scalar-Tensor Theories of Gravity}},\ }\href
  {http://ptp.oxfordjournals.org/cgi/doi/10.1143/PTP.98.359} {\bibfield
  {journal} {\bibinfo  {journal} {Prog. Theor. Phys.}\ }\textbf {\bibinfo
  {volume} {98}},\ \bibinfo {pages} {359} (\bibinfo {year} {1997})}\BibitemShut
  {NoStop}%
\bibitem [{\citenamefont {Mendes}(2015)}]{Mendes2015}%
  \BibitemOpen
  \bibfield  {author} {\bibinfo {author} {\bibfnamefont {R.~F.~P.}\
  \bibnamefont {Mendes}},\ }\bibfield  {title} {\bibinfo {title} {{Possibility
  of setting a new constraint to scalar-tensor theories}},\ }\href
  {http://link.aps.org/doi/10.1103/PhysRevD.91.064024} {\bibfield  {journal}
  {\bibinfo  {journal} {Phys. Rev. D}\ }\textbf {\bibinfo {volume} {91}},\
  \bibinfo {pages} {064024} (\bibinfo {year} {2015})}\BibitemShut {NoStop}%
\bibitem [{\citenamefont {Rosca-Mead}\ \emph {et~al.}(2020)\citenamefont
  {Rosca-Mead}, \citenamefont {Moore}, \citenamefont {Sperhake}, \citenamefont
  {Agathos},\ and\ \citenamefont {Gerosa}}]{Rosca-Mead2020}%
  \BibitemOpen
  \bibfield  {author} {\bibinfo {author} {\bibfnamefont {R.}~\bibnamefont
  {Rosca-Mead}}, \bibinfo {author} {\bibfnamefont {C.~J.}\ \bibnamefont
  {Moore}}, \bibinfo {author} {\bibfnamefont {U.}~\bibnamefont {Sperhake}},
  \bibinfo {author} {\bibfnamefont {M.}~\bibnamefont {Agathos}},\ and\ \bibinfo
  {author} {\bibfnamefont {D.}~\bibnamefont {Gerosa}},\ }\bibfield  {title}
  {\bibinfo {title} {Structure of neutron stars in massive scalar-tensor
  gravity},\ }\href {https://www.mdpi.com/2073-8994/12/9/1384} {\bibfield
  {journal} {\bibinfo  {journal} {Symmetry}\ }\textbf {\bibinfo {volume} {12}}
  (\bibinfo {year} {2020})}\BibitemShut {NoStop}%
\bibitem [{\citenamefont {Pani}\ \emph {et~al.}(2011)\citenamefont {Pani},
  \citenamefont {Cardoso}, \citenamefont {Berti}, \citenamefont {Read},\ and\
  \citenamefont {Salgado}}]{Pani2011}%
  \BibitemOpen
  \bibfield  {author} {\bibinfo {author} {\bibfnamefont {P.}~\bibnamefont
  {Pani}}, \bibinfo {author} {\bibfnamefont {V.}~\bibnamefont {Cardoso}},
  \bibinfo {author} {\bibfnamefont {E.}~\bibnamefont {Berti}}, \bibinfo
  {author} {\bibfnamefont {J.}~\bibnamefont {Read}},\ and\ \bibinfo {author}
  {\bibfnamefont {M.}~\bibnamefont {Salgado}},\ }\bibfield  {title} {\bibinfo
  {title} {{Vacuum revealed: The final state of vacuum instabilities in compact
  stars}},\ }\href {http://link.aps.org/doi/10.1103/PhysRevD.83.081501}
  {\bibfield  {journal} {\bibinfo  {journal} {Phys. Rev. D}\ }\textbf {\bibinfo
  {volume} {83}},\ \bibinfo {pages} {081501} (\bibinfo {year}
  {2011})}\BibitemShut {NoStop}%
\end{thebibliography}%

\end{document}